\documentclass[twocolumn]{aastex631}

\usepackage{xcolor}
\usepackage{verbatim}
\usepackage{amsmath}
\usepackage{hyperref}
\usepackage{tabularx}
\usepackage{multirow}

\begin{document}


\title{Probing the Faint Outskirts of Star-Forming Galaxies with $1<z<9$ Using Stacked Images}

\author[0009-0004-7042-4172]{Cheng Jia}\thanks{E-mail:  jc123@mail.ustc.edu.cn}
\affiliation{Department of Astronomy, University of Science and Technology of China, Hefei 230026, China}
\affiliation{School of Astronomy and Space Science, University of Science and Technology of China, Hefei 230026, China}

\author[0000-0003-1588-9394]{Enci Wang}\thanks{E-mail: ecwang16@ustc.edu.cn}
\affiliation{Department of Astronomy, University of Science and Technology of China, Hefei 230026, China}
\affiliation{School of Astronomy and Space Science, University of Science and Technology of China, Hefei 230026, China}

\author[0000-0002-4597-5798]{Yangyao Chen}
\affiliation{School of Astronomy and Space Science, Nanjing University, Nanjing, Jiangsu 210093, China}
\affiliation{Key Laboratory of Modern Astronomy and Astrophysics, Nanjing University, Ministry of Education, Nanjing, Jiangsu 210093, China}

\author[0009-0000-7307-6362]{Cheqiu Lyu}
\affiliation{Department of Astronomy, University of Science and Technology of China, Hefei 230026, China}
\affiliation{School of Astronomy and Space Science, University of Science and Technology of China, Hefei 230026, China}

\author{Hongxin Zhang}
\affiliation{Department of Astronomy, University of Science and Technology of China, Hefei 230026, China}
\affiliation{School of Astronomy and Space Science, University of Science and Technology of China, Hefei 230026, China}

\author{Ziwen Zhang}
\affiliation{Department of Astronomy, University of Science and Technology of China, Hefei 230026, China}
\affiliation{School of Astronomy and Space Science, University of Science and Technology of China, Hefei 230026, China}

\author{Jinyang Wang}
\affiliation{Department of Astronomy, University of Science and Technology of China, Hefei 230026, China}
\affiliation{School of Astronomy and Space Science, University of Science and Technology of China, Hefei 230026, China}

\author[0000-0002-0846-7591]{Jie Song}
\affiliation{Department of Astronomy, University of Science and Technology of China, Hefei 230026, China}
\affiliation{School of Astronomy and Space Science, University of Science and Technology of China, Hefei 230026, China}

\author[0000-0002-8420-6246]{Bojun Tao}
\affiliation{Department of Astronomy, University of Science and Technology of China, Hefei 230026, China}
\affiliation{School of Astronomy and Space Science, University of Science and Technology of China, Hefei 230026, China}

\author[0009-0006-8356-4646]{Xiaoxuan Chen}
\affiliation{Department of Astronomy, University of Science and Technology of China, Hefei 230026, China}
\affiliation{School of Astronomy and Space Science, University of Science and Technology of China, Hefei 230026, China}

\author{Junsheng Li}
\affiliation{Department of Astronomy, University of Science and Technology of China, Hefei 230026, China}
\affiliation{School of Astronomy and Space Science, University of Science and Technology of China, Hefei 230026, China}

\begin{abstract}

We present a systematic study of the faint outer light profiles of star-forming galaxies using deep JWST/NIRCam imaging from the CEERS, JADES, and PRIMER surveys. Our sample consists of 14,812 face-on disk galaxies at $1<z<9$ with $\log(M_*/M_{\odot})>8.5$, selected from the DAWN JWST Archive. To characterize the low-surface-brightness outskirts beyond the effective radius ($R_{\rm e}$), we stack galaxies in bins of stellar mass and redshift, accounting for cosmological surface-brightness dimming, angular-size scaling, and PSF convolution, and estimate residual backgrounds from blank-sky stacks. The color profiles generally become bluer toward larger radii, indicating negative color gradients and supporting an inside-out growth scenario. Galaxies with $\log(M_*/M_{\odot}) < 9.5$ at $z>2.5$ exhibit a weak reversal in their outermost regions, suggesting more complex stellar population gradients at early times. The outer light profiles generally exhibit a good exponential decline with radius, and the outer scale lengths measured beyond $2R_{\rm e}$ show mass and redshift dependences remarkably similar to those of the conventional mass–size relation. This similarity suggests that the extended outer light is closely linked to the same processes governing the growth of the inner galaxy, rather than providing a direct tracer of ex-situ stellar components in the early universe. We also find little wavelength dependence in the outer scale lengths, indicating that dust attenuation likely has a limited impact at these radii. These results provide new constraints on the buildup of extended stellar distributions during cosmic noon and at earlier epochs.

\end{abstract}


\section{introduction}\label{sec.intro}

The evolution of galaxies over cosmic time is governed by a complex interplay of physical processes acting both internally and externally. Through in-situ star formation, galaxies convert their cold gas reservoirs into new stars, driving the growth of stellar mass and contributing to structural evolution throughout their lifetimes \citep{Pillepich_2015, Chen2021, Cannarozzo_2023, Mo2024}. Ex-situ processes, particularly major and minor mergers, can substantially redistribute stars and introduce distinct stellar populations \citep{Ownsworth_2014, RG_2016, Conselice_2022, LiHao2025, Duan_2026, WangK_2026}. 
In addition, stellar migration can also drive the evolution and redistribution of stellar disks. \cite{Roskar_2008} showed that stars in the outer disk do not necessarily form in situ, but can migrate outward from the inner disk through interactions with spiral structures. 
\cite{Minchev_2012} further demonstrated that the combined effects of bars and spiral arms can enhance radial migration and redistribute angular momentum across large radial scales, potentially reshaping the radial stellar distribution of the disk outskirts.
Dynamical heating associated with the assembly of dark matter halos may also induce radial migration of pre-existing stars \citep{Chen_2024, Ma2026}. 
The relative importance of these growth channels likely varies with redshift, stellar mass, and environment \citep{Peng_2010, RG_2016, Tacchella_2019, Husko_2023, WangK_2023a, WangK_2023b, WangK_2025, Puskas_2025}, making it challenging to reconstruct the assembly histories of individual galaxies.

A particularly important challenge is to reconcile the observed size growth of high-redshift galaxies with their spatially resolved stellar population properties. For low-mass galaxies at $z>2$, recent studies find enhanced specific star formation rates (sSFRs) in their central regions \citep{Tripodi_2024, Jia_2024, Song_2026, Lyu_2025}. Such centrally concentrated in-situ star formation would preferentially build the inner galaxy and thus tend to suppress or even reverse size growth \citep{Chen2025, Ma2026}, in apparent tension with the rapid size evolution observed at similar epochs \citep{Ormerod_2023, Ward_2024, Morishita_2024, Jia_2024, Yang_2025, Allen_2025}. This tension suggests that in-situ star formation alone may not fully account for the rapid growth of galaxy sizes in the early Universe, and that ex-situ accretion and/or radial redistribution of pre-existing stars may play an important role \citep{Cole_2002, Naab_2009, WangK_2026}.
It is therefore crucial to develop observational approaches that can disentangle and quantify the relative contributions of different growth channels in high-redshift galaxies. In particular, the faint outer regions, where signatures of ex-situ accretion and stellar redistribution are expected to be most prominent, provide a promising window into the mechanisms driving galaxy assembly.

This motivates the search for observational signatures that can distinguish the different channels of stellar mass assembly. The stellar halo—the diffuse, low-surface-brightness component extending well beyond the main body of a galaxy—offers a unique window into the external assembly history of galaxies \citep{Bullock_2005, Helmi_2020, Pulsoni_2021, Zhu_2022, Gozman_2023}. Unlike the bright inner regions, where ongoing star formation, dynamical relaxation, and secular processes can continuously reshape stellar populations and obscure their assembly histories, the outskirts are expected to preserve signatures of stellar material deposited through past mergers and accretion events \citep{Elias_2018, Di-Matteo_2019, Merritt_2020, Wright_2024, Gonzalez-Jara_2025}. The stellar halo therefore provides a potential fossil record of galaxy assembly, offering an avenue for separating the contributions of in-situ star formation and ex-situ accretion to stellar mass growth. Characterizing the faint outer regions across cosmic time is thus essential for understanding the relative importance of these growth channels and how they shape galaxy structure.

The properties and origins of stellar halos have been extensively investigated at relative low redshift from both observational and theoretical perspectives \citep{Morrison_2000, Bullock_2005, Bell_2008, Tumlinson_2009, WangWT_2019, Pulsoni_2021, Keller_2022, Dodd_2025, Ogami_2025, Xu_2025, Cooper_2025, Li_2026, Tao_2026, Suzuki_2026}. 
The outer light profiles of galaxies commonly exhibit up-bending or down-bending breaks in both the local and distant Universe \citep{Erwin_2008, Borlaff_2018}. However, \cite{Bakos_2008} found that these breaks, which are prominent in light profiles, can be substantially weaker in stellar mass surface-density profiles.
Deep imaging combined with stacking techniques has enabled measurements of galaxy outskirts over scales of several hundred kiloparsecs. 
\cite{WangWT_2019} demonstrated the power of stacking HSC imaging by tracing the surface brightness profiles of isolated central galaxies down to $\sim31~{\rm mag~arcsec^{-2}}$. They found that more massive galaxies host redder and more extended stellar halos, with the outer profiles exhibiting strong self-similarity when scaled by the halo virial radius.
More recently, \cite{Xu_2024} and \cite{Yu_2025} investigated disk breaks in massive galaxies at $1 < z < 3$ and found that a fraction of ``down-bending'' disks exhibit U-shaped color profiles.
\cite{Tao_2026}, using HSC-SSP deep imaging, conducted a systematic census of stellar halos in 169 high-inclination central galaxies and measured their stellar halo fractions. 
\cite{Adams_2026} conducted a statistical study of low-surface-brightness tidal structures in more than 19,000 edge-on galaxies, finding that such features occur in only $\sim 5–6\%$ of systems.
\cite{Williams_2025} extended direct detections of larger scale stellar halos to $z\sim1$ using HST imaging. On the theoretical side, simulations suggest that the outer regions of galaxies can provide sensitive constraints on feedback models \citep{Keller_2022} and gas-accretion geometry \citep{Ma2026, Ma2026b, Lyu2026}, while several studies have proposed that outer light can serve as a proxy for the ex-situ stellar component \citep{Elias_2018, Merritt_2020}. 
Although some studies of the outer regions of galaxies during cosmic noon have been carried out \citep{Xu_2024, Yu_2025}, more systematic investigations spanning a wider range of redshifts and stellar masses remain scarce.

The advent of the James Webb Space Telescope (JWST) has opened a new window for investigating the low-surface-brightness structures of high-redshift galaxies. Its high spatial resolution and unprecedented near-infrared sensitivity enable observations that are substantially deeper and more spatially resolved than those available from previous facilities, particularly at rest-frame optical wavelengths for galaxies in the early Universe. Although existing JWST extragalactic surveys have generally prioritized ultra-deep observations over wide-area coverage, their exceptional depth and broad wavelength coverage provide a unique opportunity to statistically probe the faint outskirts of intermediate- and high-redshift galaxies. In particular, JWST imaging allows us to investigate the rest-frame optical light distributions of galaxies during cosmic noon and cosmic dawn, providing new constraints on the spatial distribution of their older stellar populations and stellar mass.

In this work, we exploit deep JWST/NIRCam imaging to systematically investigate the photometric properties of the outskirts of star-forming galaxies over $1<z<9$. 
Using stacked images, we focus on radii of approximately $2$--$4R_{\rm e}$, a transitional region between the main galactic body and the more extended stellar halo \citep{Cook_2016, Williams_2025}. 
At these redshifts, this radial range typically corresponds to physical scales of $\sim10$ kpc. Although these regions do not yet reach the conventionally defined stellar-halo regime—the limiting physical scale of our measurements is approximately $15$ kpc—they provide a critical and largely unexplored interface between the inner galaxy and its diffuse outer component. We characterize the surface-brightness and color profiles of these outskirts across a broad range of stellar masses and redshifts, investigate their connection to the inner galaxy, and examine the evolution of their spatial structure. 
Our goal is not to directly identify stellar halos at high redshift, but rather to establish observational constraints on the faint outer light profiles, which may arise from a variety of stellar components, including stellar halos and stars redistributed from the inner regions through gravitational instabilities and spiral/bar scattering, thereby providing a foundation for future studies of the origin and evolution of galaxy outskirts in the early Universe.


This paper is organized as follows. In Section \ref{sec.data}, we describe the sample selection, image-stacking methodology, and background estimation. In Section \ref{sec.res}, we present our results, including the surface brightness profiles, color gradients, radial extent of the galaxy outskirts, and their evolution with redshift and stellar mass. In Section \ref{sec.sum}, we summarize our main findings and discuss their implications. Throughout this paper, we adopt a flat $\Lambda$CDM cosmology with $\Omega_{\rm m}=0.308$, $\Omega_{\Lambda}=0.691$, and $H_0=67.74~\mathrm{km~s^{-1}~Mpc^{-1}}$ \citep{Planck_2016}.

\section{data and method}\label{sec.data}

\subsection{sample definition}

We combine JWST/NIRCam imaging from the JADES \citep{jades_2023}, CEERS \citep{ceers_2025}, and PRIMER \citep{primer_2021} survey fields using the mosaics provided by the DAWN JWST Archive (DJA) Mosaic Release Version 7. These mosaics are processed with the {\tt grizli} pipeline \citep{Valentino_2023}. Our dataset includes seven fields: GDS (v7.2, $\sim72.8~{\rm arcmin^2}$), GDN (v7.3, $\sim94.4~{\rm arcmin^2}$), CEERS (v7.2, $\sim105.1~{\rm arcmin^2}$), UDS-S (v7.2, $\sim144.3~{\rm arcmin^2}$), UDS-N (v7.2, $\sim149.3~{\rm arcmin^2}$), COSMOS-E (v7.0, $\sim106.8~{\rm arcmin^2}$), and COSMOS-W (v7.0, $\sim102.1~{\rm arcmin^2}$). Together, these fields provide a total area of approximately $0.2~{\rm deg^2}$. The fields have been observed with both HST/WFC3 and JWST/NIRCam, providing imaging over a wavelength range of approximately $0.4$--$4~\mu{\rm m}$.

For morphological analysis and the estimation of physical properties, we use the HST F435W, F606W, and F814W bands, together with the JWST/NIRCam F090W (except in CEERS), F115W, F150W, F200W, F277W, F356W, and F444W bands. For the image-stacking analysis, we use only the JWST/NIRCam bands. 
We adopt the SourceExtractor++ Morphological Catalogs from DJA\footnote{\url{https://dawn-cph.github.io/dja/blog/2024/08/16/morphological-data/}} as our parent catalog. 
Specifically, we use the redshift, coordinates and multi-band photometry from these catalogs.
To remove faint sources and objects with unreliable photometric redshift estimates, we apply the same selection criteria as those described in Section 2.3 of \cite{Song_2026}.
We require a $S/N>10$ in the detection image and $S/N>3$ in each of the F115W, F150W, F200W, F277W, F356W, and F444W bands. To ensure the reliability of the photometric redshift estimates, we further require $\chi^2/N_{\rm filter}\leq8$ and $N_{\rm filter}\geq6$ in the photometric redshift fitting. In addition, we exclude sources with $m_{\rm F444W}>28.5$ mag.
It yields 206,790 sources with reliable photometric measurements.

We first identify star-forming galaxies using the criterion ${\rm sSFR}>1/3t_{\rm H}$ \citep{Bluck_2023}, where $t_{\rm H}$ is the Hubble time of corresponding redshift. We then apply a series of morphological criteria to construct a clean sample of galaxies suitable for measuring their faint outer light profiles. We derive the structural parameters using {\sc Galfit} \citep{Peng_2002, Peng_2010} and select face-on disk galaxies with a minor-to-major axis ratio of $b/a>0.5$, where $b/a$ is interpolated to the rest-frame $1 \ \mu\mathrm{m}$. If $1 \ \mu\mathrm{m}$ is not covered by the available bands, we adopt the $b/a$ measured in the band closest to the rest-frame $1 \ \mu\mathrm{m}$.
To ensure reliable structural measurements, we exclude galaxies for which the difference between the {\sc Galfit} and DJA catalog magnitudes exceeds 0.5 mag, as well as galaxies whose fitted radii are substantially larger than the corresponding segmentation-map sizes. We further remove potential major merger systems by requiring an F444W-band asymmetry parameter $A<0.35$, measured with the {\tt statmorph\_csst} package \citep{Yao_2023}. These selections yield 46,657 galaxies.

\begin{figure}[ht]
    \centering
    \includegraphics[scale=0.42]{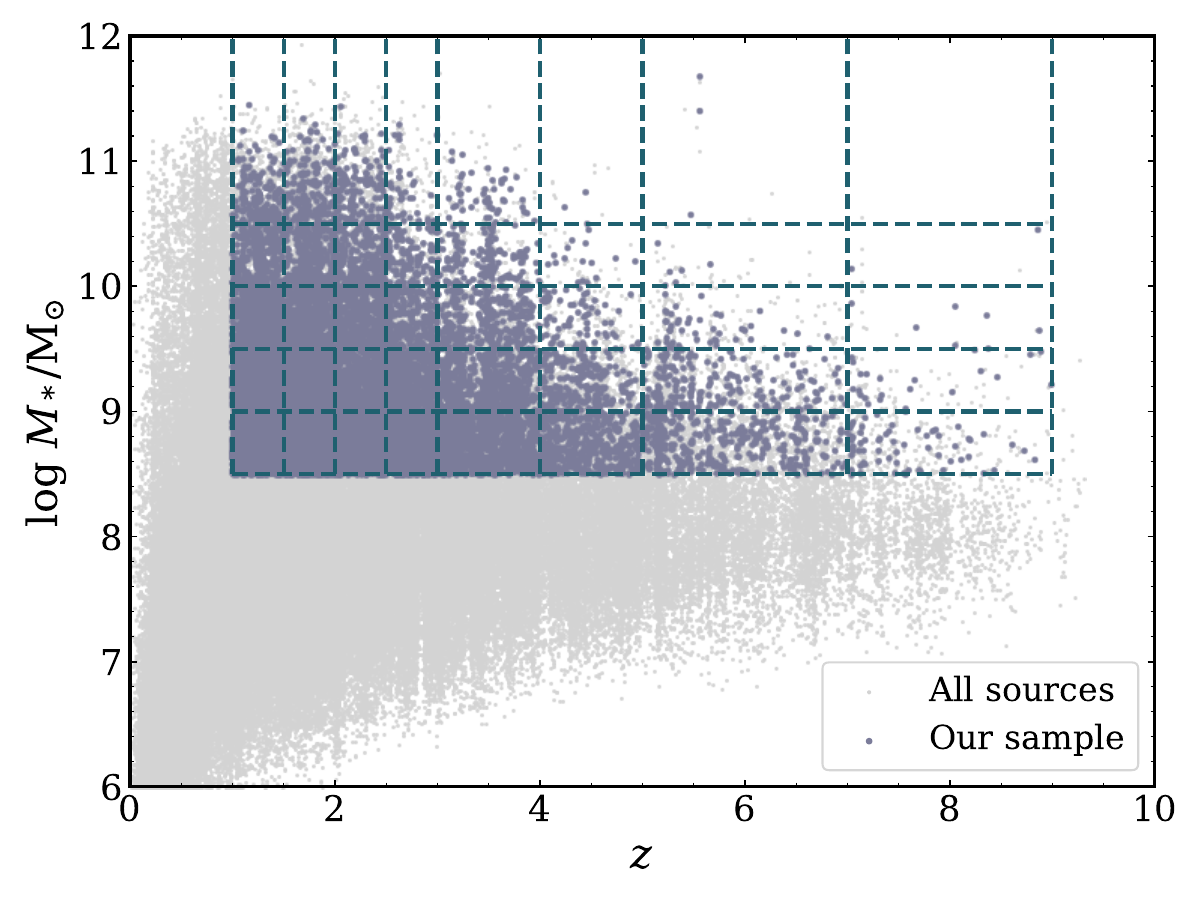}
    \caption{
    Sample selection in the stellar mass--redshift plane. Light-gray points show the parent sample, while purple points indicate the final selected sample. Dark-green dashed lines mark the adopted binning scheme.
    }
    \label{Fig.sample}
\end{figure}

We derive the physical properties of the selected galaxies, including their stellar masses and SFRs, using {\tt CIGALE}, following the procedure described in \cite{Song_2026}. The stellar masses are used to divide the galaxies into stellar-mass bins, while the SFRs are used to identify and exclude quiescent galaxies, as described above.
We adopt spectroscopic redshifts when available and photometric redshifts otherwise, with both measurements taken from the DJA catalog. Finally, to ensure reliable image quality while maintaining sufficient wavelength coverage and a broad redshift baseline, we restrict the sample to galaxies with $\log(M_*/{\rm M_{\odot}})>8.5$ and $1<z<9$. The resulting final sample contains 14,812 star-forming, face-on disk galaxies. The distributions of the final sample in redshift and stellar mass are shown in Figure \ref{Fig.sample}.

\subsection{stacking process}


Given the absence of F090W imaging in the CEERS field, we restrict our stacking analysis to the JWST/NIRCam bands from F115W to F444W. To ensure consistent spatial resolution across wavelengths, we first convolve all galaxy images to a common PSF corresponding to the F444W band, as shown in \cite{Song_2026}. 
For each galaxy, we extract a $\sim200\times200$ pixel cutout, which is sufficiently large to encompass the radial range over which the stacked surface-brightness profiles remain above $3\sigma_{\rm bkg}$, where $\sigma_{\rm bkg}$ is the background noise.
We divide the sample into redshift and stellar mass bins, as illustrated in Figure \ref{Fig.sample}, and stack the galaxy images in each bin.
Because our sample spans a broad redshift range and individual redshift bins cover substantial intervals, galaxies at different redshifts would otherwise be subject to different levels of cosmological surface-brightness dimming and angular-size scaling, potentially biasing the stacked profiles and complicating comparisons between redshift bins. We therefore transform all images to a common reference redshift of $z=1.5$ before stacking. First, we correct each image for cosmological surface-brightness dimming. Since the mosaics are calibrated in nJy, we multiply the flux density of each pixel by $(1+z)^3/2.5^3$, effectively removing the $(1+z)^3$ dimming relative to the reference redshift of $z=1.5$ \citep{Lerner_2014, Williams_2025}. 
We then rescale the angular size of each galaxy according to the ratio of the angular-diameter distance at its redshift to that at $z=1.5$, thereby placing galaxies at different redshifts on a common physical scale before stacking. The resampling is performed using the \texttt{scipy.ndimage.zoom} function, which interpolates the images to the desired pixel scale. To ensure consistent PSF treatment, we also rescale the PSF for each redshift bin using the central redshift of that bin, applying the same rescaling procedure as for the images. This ensures that the PSF pixel scale is consistent with that of the corresponding resampled images. The rescaled PSFs are then used in the subsequent {\sc Galfit} fitting process.

\begin{figure*}[ht]
    \centering
    \includegraphics[scale=0.33]{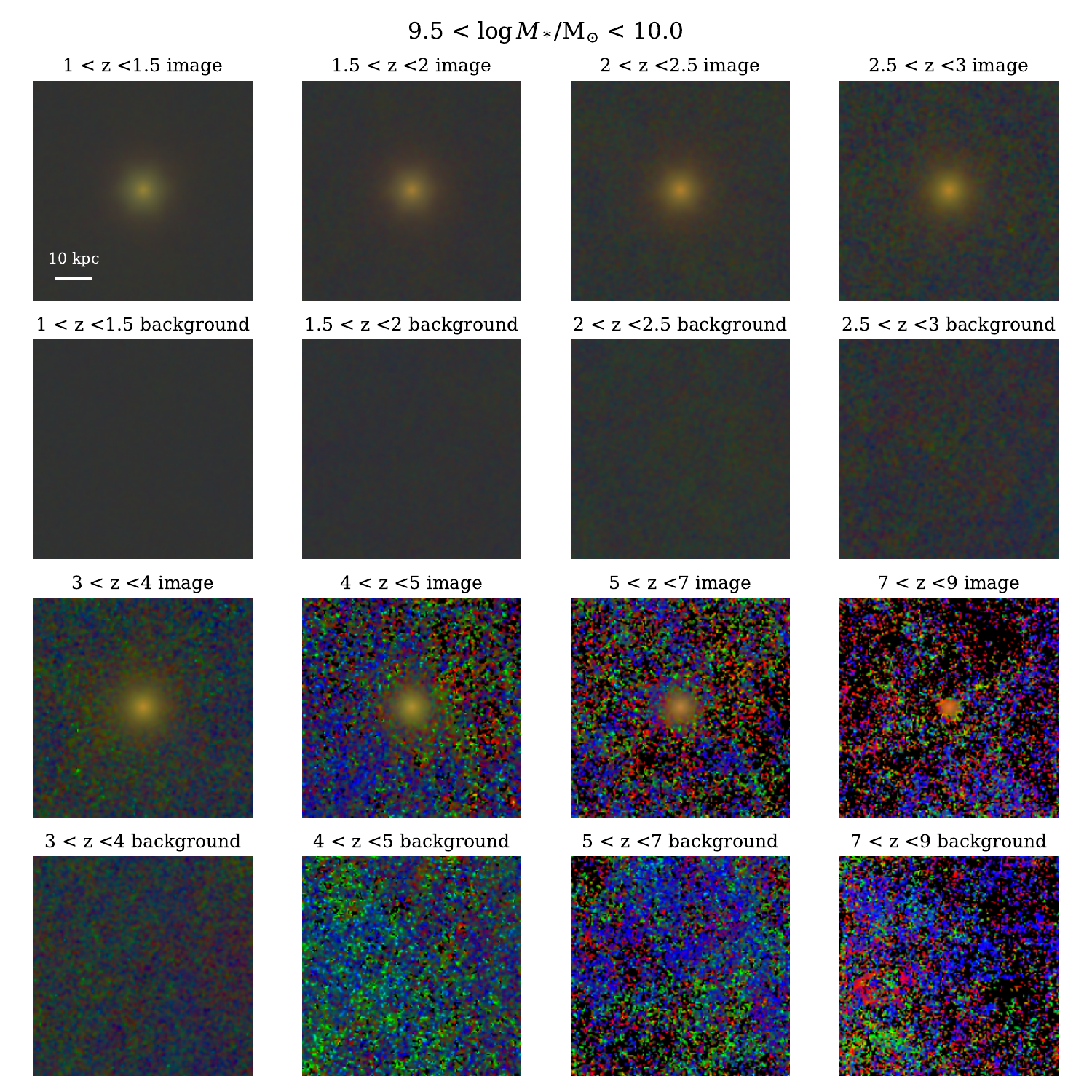}
    \includegraphics[scale=0.33]{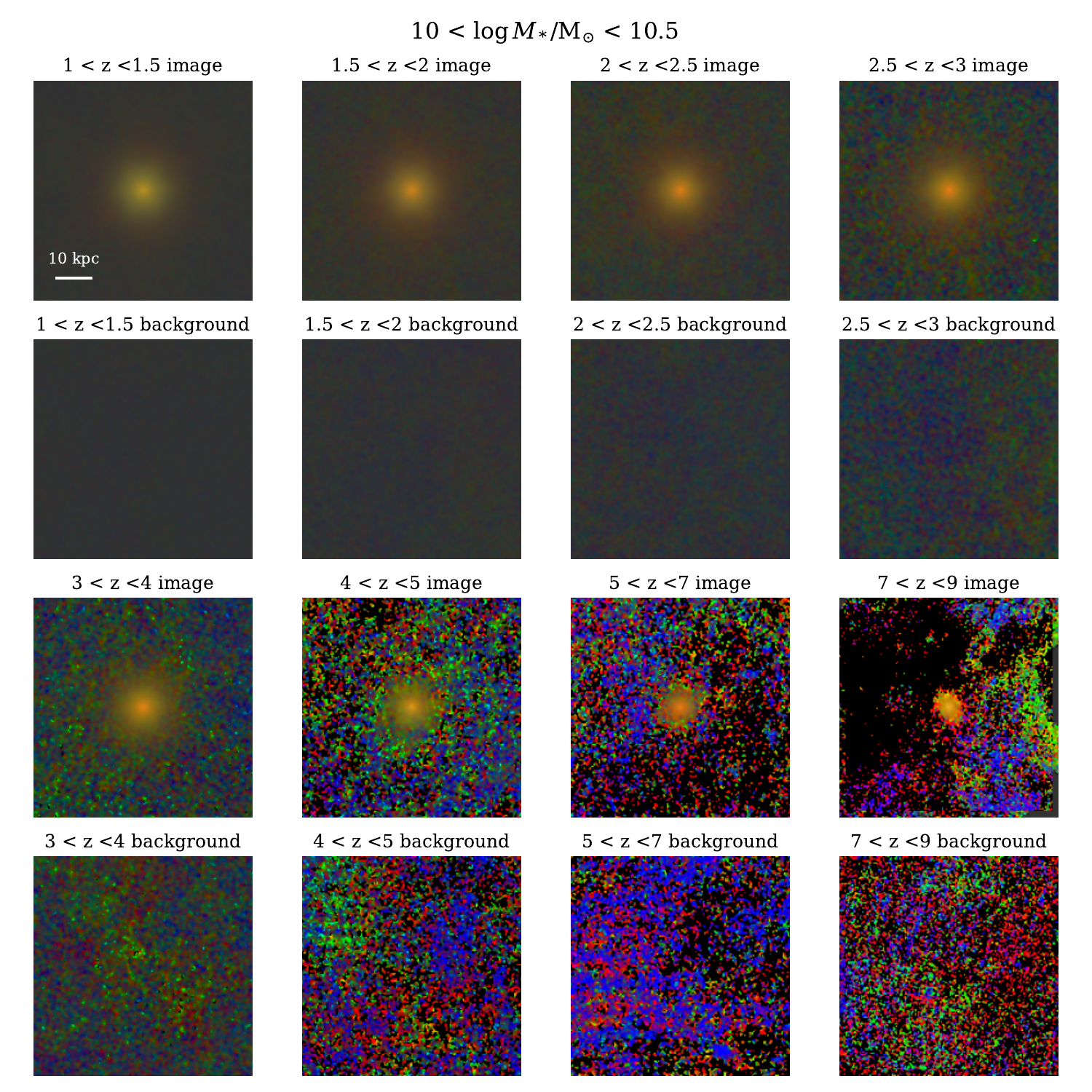}
    \caption{Examples of stacked galaxy images and the corresponding background estimates. The left four columns show stacks for galaxies with $9.5 < \log M_*/{\rm M_{\odot}} < 10$, while the right four columns show stacks for $10 < \log M_*/{\rm M_{\odot}} < 10.5$. Owing to space constraints, stacks for the other stellar-mass bins are presented in the Appendix \ref{app.imgs}. The multi-band color composites are constructed from the F115W, F277W, and F444W images. We adopt a consistent stretch of 1 for the RGB images of both the stacked images and the background images at all redshifts. For ease of comparison, all images are rescaled to a common reference redshift of $z=1.5$, as described in Section~\ref{sec.data}.
    }
    \label{Fig.examples}
\end{figure*}

To minimize contamination from neighboring sources, we mask pixels associated with other detected sources using the corresponding segmentation maps. The masked pixels are set to {\tt NaN} and excluded from the stacking process, and therefore these regions do not affect the stacked images.
After applying the corrections described above, all pixels in a given redshift and stellar-mass bin are placed on a common physical scale are subject to the same cosmic dimming effect.
We then construct the stacked image by calculating the mean value of corresponding pixels across all galaxy stamps within each bin. To reduce the impact of extreme pixel values, we perform a $10\%$ two-sided clipping, removing the upper and lower $10\%$ of the pixel-value distribution before calculating the mean \citep{WangWT_2019}. Because the galaxies are not rotated prior to stacking and their position angles are randomly distributed, the resulting stacked images are effectively azimuthally symmetric. Examples of the stacked images for galaxies with $9.5 < \log M_*/{\rm M_{\odot}} < 10$ and $10 < \log M_*/{\rm M_{\odot}} < 10.5$ are presented in Figure \ref{Fig.examples}.

The presence of the PSF can significantly affect measurements of the faint outer light profiles, particularly at small radii. To quantify and correct for this effect, we use {\sc Galfit} to model the stacked images. We construct the corresponding error maps through bootstrap resampling. For each bin, we randomly draw the individual galaxy stamps with replacement for 100 iterations and calculate the standard deviation of the resulting pixel values at each position, which we adopt as the pixel-level uncertainty. Considering that the PSF orientation varies slightly among the different fields and that the stacked images are axisymmetric, we adopt the PSF of the GDS field as representative of the sample and do not apply an additional correction for PSF orientation. For the same reason, we fix the center to the center of stacked image, the position angle to 0, and the axis ratio ($b/a$) to 1 in the {\sc Galfit} fits, allowing only the total magnitude, effective radius, and S\'ersic index to vary. Given the limited spatial resolution of high-redshift galaxies, we find that a single S\'ersic profile provides an adequate description of the majority of the stacked images while keeping the fitted parameters physically interpretable. We therefore adopt a single S\'ersic model for all stacked images. Using the best-fit S\'ersic parameters, we construct a PSF-free model image and add the fitting residuals back to the model as a first-order correction. The resulting image provides an estimate of the intrinsic, PSF-corrected light distribution, which we use for the subsequent analysis.
However, we note that this deconvolution procedure may be unreliable for galaxies whose effective radii are comparable to the PSF size \citep{Miller2021, Xu_2024}. We therefore show the corresponding PSF FWHM in the F444W band for each redshift bin in Figure \ref{Fig.profile}, allowing a direct comparison between galaxy sizes and the spatial resolution of the observations.


When studying the faint outskirts of galaxies, accurate background characterization is critical, particularly for bins containing relatively small numbers of galaxies, such as those at high redshift and high stellar mass. We therefore estimate the background independently for each redshift--stellar-mass bin using the following procedure.
For each galaxy in a given bin, we randomly select a blank-sky region from the same field and extract a cutout with the same size as the corresponding galaxy stamp. These blank-sky cutouts are then processed using exactly the same procedure as the galaxy images, including the cosmological surface-brightness dimming correction, angular-size rescaling, and pixel-by-pixel $10\%$ two-sided clipping. We subsequently stack the processed blank-sky cutouts using the same procedure adopted for the galaxy images. This yields a background map for each stacked image, examples of which are shown in Figure \ref{Fig.examples}. As shown in Figure \ref{Fig.examples}, the amplitudes of the estimated background maps are generally comparable to the noise level of the stacked images, indicating that the residual background is small and does not dominate the measured outer light profiles.

\section{results}\label{sec.res}

\subsection{stacked surface brightness profiles}\label{subsec.prof}


The stacked surface brightness profiles in the F444W band are presented in Figure \ref{Fig.profile}, while the profiles in the other bands are shown in Appendix \ref{app.prof}. We include only redshift--stellar-mass bins containing at least 10 galaxies to ensure reliable measurements. For each profile, we define the limiting detection radius as the outermost radius at which the surface brightness remains above the background threshold, $\Sigma_{\rm bkg}+3\sigma_{\rm bkg}$, where $\Sigma_{\rm bkg}$ is the median surface brightness of the stacked background image and $\sigma_{\rm bkg}$ is its standard deviation, namely, the background noise. The profiles are truncated at this limiting detection radius.

\begin{figure*}[htbp]
  \centering
  \includegraphics[width=0.95\textwidth]{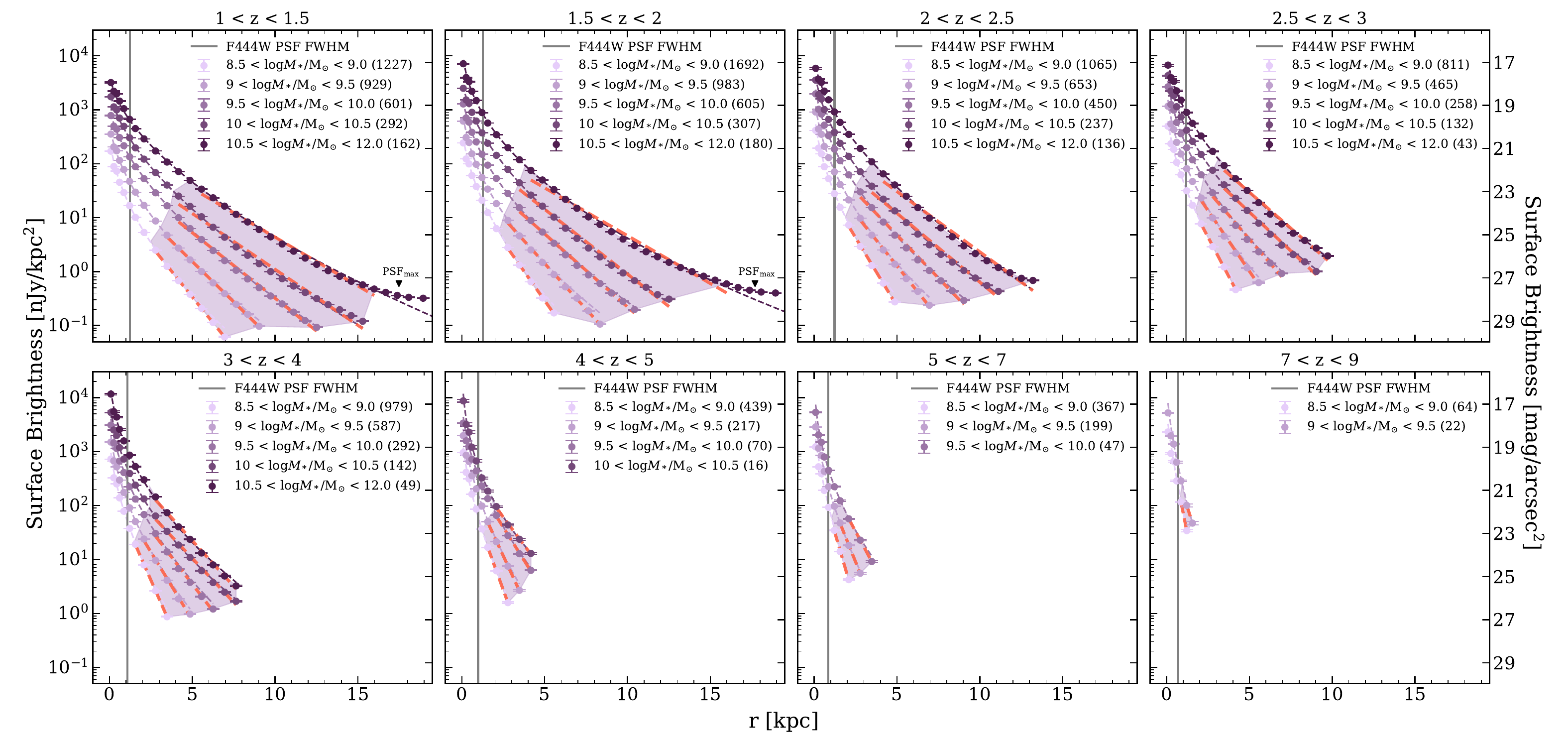}
  \caption{
  Surface brightness profiles derived from the stacked images in the F444W band for illustration. 
  The fluxes and errors are both measured using the {\tt photutils.aperture} package through aperture photometry in circular annuli at each radius.
  We emphasize that all profiles are measured from the stacked images rescaled to the common reference redshift of $z=1.5$.
  For each profile, the corresponding stellar-mass bin and number of stacked galaxies are indicated in the legend.
  The dotted lines with corresponding colors are the best-fit S\'ersic model derived from {\sc Galfit}.
  Red dotted lines are the best-fit exponential model in Section \ref{sec.hR}. The $\rm PSF_{max}$ indicates the truncation radius of the ePSF image.
  The shaded regions indicate the radial ranges between $2R_{\rm e}$ and the adopted outer fitting limits, with a maximum radius of 16 kpc for the most massive galaxies at $z<2$. These regions are used to measure the outer scale lengths in Section \ref{sec.hR}. Surface brightness profiles in the other bands are presented in Appendix \ref{app.prof}.
  }
  \label{Fig.profile}
\end{figure*}

As shown, the stacked surface brightness profiles remain smooth and well detected to radii beyond $\sim5-10$ kpc. For the most massive galaxies at $z\sim1.5$, the outer light can be traced to $\sim20$ kpc at the $3\sigma$ level, demonstrating that our stacking technique enables us to probe the faint outskirts of high-redshift galaxies to unprecedented depths.

At all redshifts, more massive galaxies tend to exhibit more extended outer light profiles, consistent with previous studies at lower redshifts \citep{WangWT_2019, Williams_2025}. At fixed stellar mass, the detectable extent of the outer light decreases toward higher redshift, with the limiting detection radius shrinking by on average a factor of four from $z\sim1.5$ to $z\sim4$. For galaxies at $z>7$, the detectable signal is generally limited to radii of $\sim1$ kpc. Several factors contribute to this rapid decline in the detectable extent. First, the substantially smaller number of galaxies available at high redshift, combined with the stronger cosmological surface-brightness dimming, results in lower signal-to-noise ratios in the stacked images. Second, galaxies at fixed stellar mass were intrinsically more compact in the early Universe \citep{van_der_Wel_2014, van_Dokkum_2015, Mowla_2019, Nedkova_2021, Ward_2023, Jia_2024, Song_2026}, naturally leading to less extended outer light profiles. We investigate the evolution of the outer scale lengths in more detail in Section \ref{sec.hR}.

We caution that a direct comparison of the surface brightness profiles measured in the same observed band across different redshifts is not physically straightforward, because a fixed observed wavelength corresponds to substantially different rest-frame wavelengths over our broad redshift range. We therefore focus primarily on the relative trends within each redshift interval and defer a detailed discussion of the wavelength dependence to later sections.


We find that for the vast majority of stacked profiles, the outskirts of galaxies do not show any significant excess or break relative to the outward extrapolation of the inner disk profiles. 
However, we emphasize that this smooth feature may arise from the stacking process. \cite{Xu_2024} found that, for individual galaxies at redshifts of $1<z<3$, the outskirts can exhibit either up-bending or down-bending profiles, with these two types accounting for 34.8\% and 56.7\% of the total sample, respectively. The varying radii at which these breaks occur, together with the inclusion of galaxies that are not perfectly face-on in the stacking process, may smooth out the breaks in stacked profiles. 
Nevertheless, this result indicates that, statistically, the surface brightness of galaxy outskirts declines smoothly, without a systematic up-bending or down-bending trend.
An interesting feature in Figure \ref{Fig.profile} is that the surface brightness profiles of massive galaxies at $z<2$ exhibit a noticeable flattening at radii of $\sim16$ kpc. This feature is detected only in F277W and redder bands, which at these redshifts correspond approximately to rest-frame wavelengths around $1~\mu{\rm m}$. In the three bluer bands, the profiles at similar radii have already fallen below the background limit described above. A similar break or flattening in the outer surface brightness profiles has also been reported at $z < 1$ \citep{Williams_2025} and is generally interpreted as the profiles approaching the surface brightness limit, beyond which the measurements become dominated by background fluctuations.

This interpretation is certainly possible for our measurements. However, we have adopted a conservative $3\sigma$ threshold based on the reconstructed background maps, and we find that subtracting the median background level from the images has a negligible impact on the results, as the average background level is very low. This suggests that the observed flattening is unlikely to be caused by the background.
We find that the flattening observed in the outskirts of these profiles may be attributed to the finite extent of the ePSF images used in our analysis, as indicated by $\rm PSF_{max}$ in Figure \ref{Fig.profile} (and also in Figure \ref{Fig.profile_115}-\ref{Fig.profile_356}). Given the limited number of reliable bright stars available in each field, it is not feasible to construct reliable PSFs extending sufficiently far into the outskirts. However, this effect is relevant only when the profiles extend to radii approaching the ePSF truncation radius. For the vast majority of redshift and stellar-mass bins, the profiles are instead truncated at smaller radii by our conservative $3\sigma$ background threshold, estimated from the reconstructed background maps. To prevent the few cases where the profiles extend close to the ePSF truncation radius from affecting our subsequent analysis, we exclude the light beyond $\sim16$ kpc from all subsequent quantitative analyses.




\subsection{color profiles}

\begin{figure*}[htbp]
  \centering
  \includegraphics[width=1.0\textwidth]{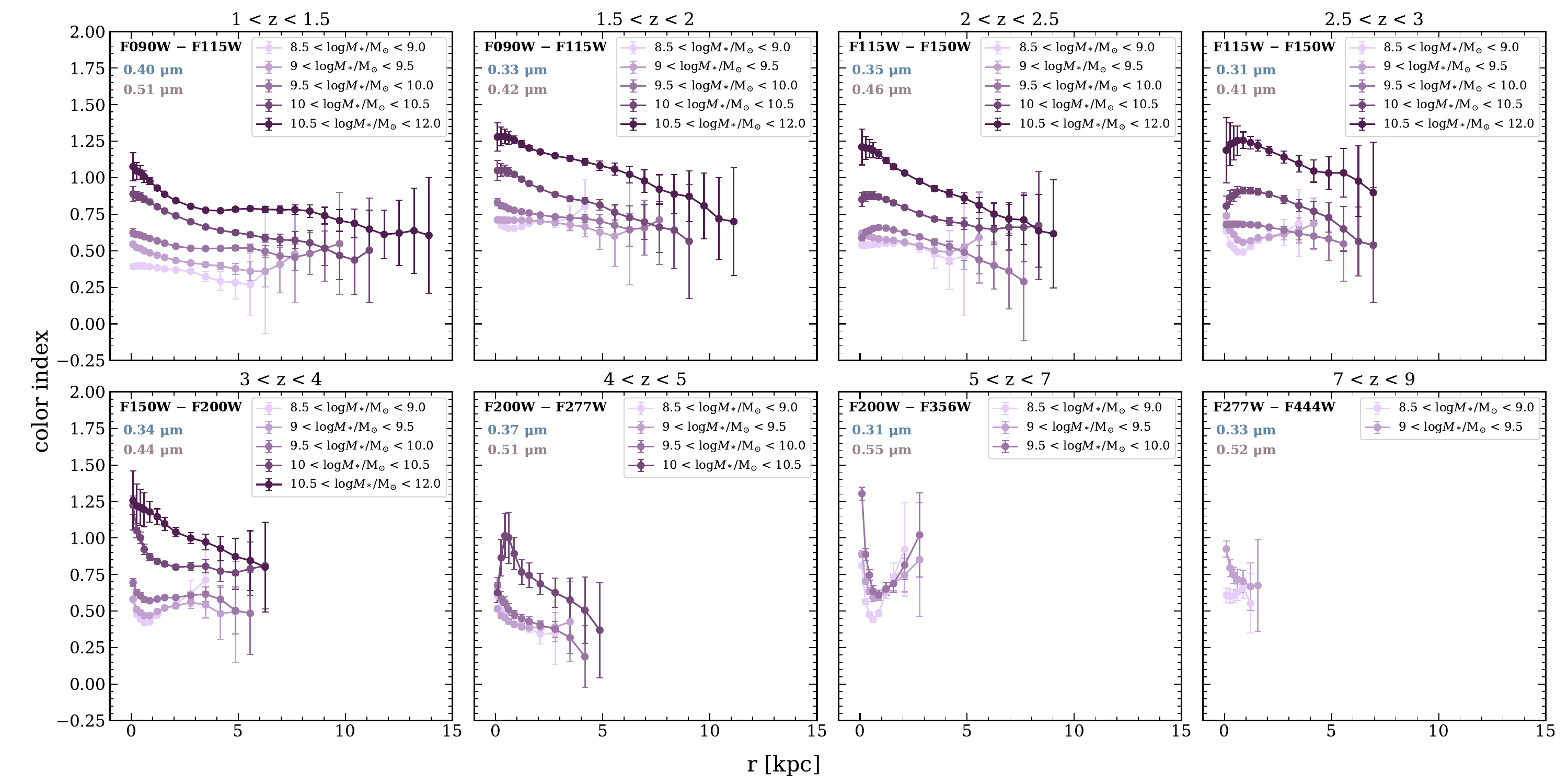}
  \caption{
  Color profiles for galaxies in different stellar-mass and redshift bins. For the two lowest-redshift bins, we use the F090W band, which is not available in the CEERS field. The corresponding rest-frame wavelength ranges are labeled in the upper-left corner of each redshift panel.
  }
  \label{Fig.color}
\end{figure*}

The PSF correction process enables us to measure the intrinsic radial color profiles of galaxies.
However, given our broad redshift range, it is not possible to adopt a single pair of observed bands that corresponds to the same rest-frame wavelength interval across the entire sample. To maximize the physical comparability of the colors between redshift bins and to probe stellar population properties, we therefore select, for each redshift bin, the pair of observed bands whose effective rest-frame wavelengths most closely bracket the $4000~\mathrm{\AA}$ break. We then measure the color from these two bands after subtracting the median background level independently for each band and redshift bin.

The strength of the $4000~\mathrm{\AA}$ break is primarily sensitive to the age of the underlying stellar population, with older populations generally exhibiting a stronger break \citep{Kauffmann_2003, Kriek_2006, Li2015, Wang2018, Wang2020}. Colors constructed from bands that bracket this feature therefore provide a useful tracer of the relative mean stellar age and its radial variation. We caution, however, that these color measurements can also be affected by dust attenuation, and should therefore not be interpreted as a pure age indicator.

Because of the limited wavelength coverage, we use the F090W stacked images to construct the colors in the two lowest-redshift bins. Since F090W imaging is unavailable in the CEERS field, we construct the F090W stacks using only galaxies from the JADES and PRIMER fields. For the other bands, we include galaxies from all fields to maximize the signal-to-noise ratio. As demonstrated in Appendix \ref{app.ceers}, excluding CEERS from the F090W stacking does not introduce significant systematic differences, indicating that the resulting color profiles are robust.

The color profiles of the stacked galaxies are presented in Figure \ref{Fig.color}. The selected filters and rest‑frame wavelengths corresponding to the observed bands used at each redshift are labeled in the upper‑left corner of each panel.

Overall, galaxies with higher stellar masses tend to exhibit redder colors at all redshifts. At $z<2.5$, galaxies across the full stellar-mass range generally show a red-to-blue color gradient from the center outward, with the gradient becoming progressively steeper toward higher redshift. This trend is broadly consistent with an inside-out growth scenario, in which star formation is more centrally concentrated at earlier times. However, centrally enhanced dust attenuation, particularly in massive galaxies, can produce a similar color gradient and may therefore contribute to the observed trend \citep{Hodge_2016, Barro_2016, G_mez_Guijarro_2022, Li_2026}.

An interesting exception is found for the most massive galaxies ($\log M_*/{\rm M_{\odot}}>10.5$) at $1<z<1.5$. Their colors decrease rapidly from the center to $\sim3$ kpc, approximately corresponding to the effective radius ($R_{\rm e}$) of the stacked image, but remain nearly constant over $\sim3$--$10$ kpc. This radial range is still largely contained within the galactic disk. The nearly flat color profile may indicate relatively homogeneous stellar populations across the outer disk, potentially reflecting a more sustained mode of gas accretion and star formation in massive galaxies across the disk at these relatively low redshifts.

At higher redshifts, the overall trend of more massive galaxies being redder persists, but the radial color profiles become more diverse. Massive galaxies generally retain the red-center/blue-outskirts configuration, whereas galaxies with $\log M_*/{\rm M_{\odot}}<9.5$ tend to show relatively bluer central regions than their outskirts.  This behavior is unlikely to be driven primarily by dust attenuation, as the central regions of galaxies are generally expected to have higher gas column densities and, consequently, stronger dust attenuation. Instead, it may be related to the transition in galaxy growth mode around cosmic noon reported by \cite{Song_2026}, who found from spatially resolved SED fitting that galaxies at high redshift preferentially undergo outside-in growth \citep{Lyu_2025}, as indicated by their radial sSFR profiles.

In the intermediate-redshift range of $2.5<z<4$, low-mass galaxies exhibit a tentative ``checkmark''-shaped color profile, becoming bluer from the center outward before turning redder again in the outermost regions. Such an upturn may reflect the combined effects of stellar population age and dust attenuation across the galaxy. For example, a radially declining dust attenuation together with an increasing stellar population age toward the outskirts could contribute to the observed ``checkmark''-shaped color profile. However, we emphasize that this feature should be interpreted with caution, as a detailed analysis of the degeneracy between dust attenuation and stellar population properties is required to verify its physical origin.
In addition, the colors of galaxies with different stellar masses tend to converge at large radii, where the surface brightness approaches the detection limit. 
The ``checkmark''-shaped feature could also be caused by imperfections in the PSF correction, particularly when the effective radius is comparable to the PSF FWHM \citep{Miller2021, Xu_2024}.
Given the relatively large uncertainties in these outer regions, this apparent convergence may be affected by uncertainties in background subtraction and should therefore be interpreted with caution.



\subsection{scale length of galaxy outskirts}\label{sec.hR}


As discussed above, the limiting detection radius of the stacked images depends not only on the intrinsic extent of the outer stellar light, but also on cosmological surface-brightness dimming and the number of galaxies contributing to each stack. In bins with relatively few galaxies, for example, the outer surface brightness can rapidly fall below the $3\sigma$ background threshold, causing the measured profile to be truncated. The limiting detection radius therefore does not provide a robust measure of the intrinsic extent of the outer light across different galaxy populations.

To characterize the outer light distribution more robustly, we instead use the scale length of the outer surface brightness profile, defined as the radial interval over which the surface brightness decreases by a factor of $e$. This is motivated by the fact that the outer light profiles beyond $2R_{\rm e}$ are well described by exponential functions, as shown in Figure \ref{Fig.profile}.
Provided that the scale length is measured within a region with sufficiently high signal-to-noise ratio, it is less sensitive to the background limit and therefore provides a more reliable measure of the spatial extent of the outer light.

\begin{figure*}[htbp]
  \centering
  \includegraphics[width=0.95\textwidth]{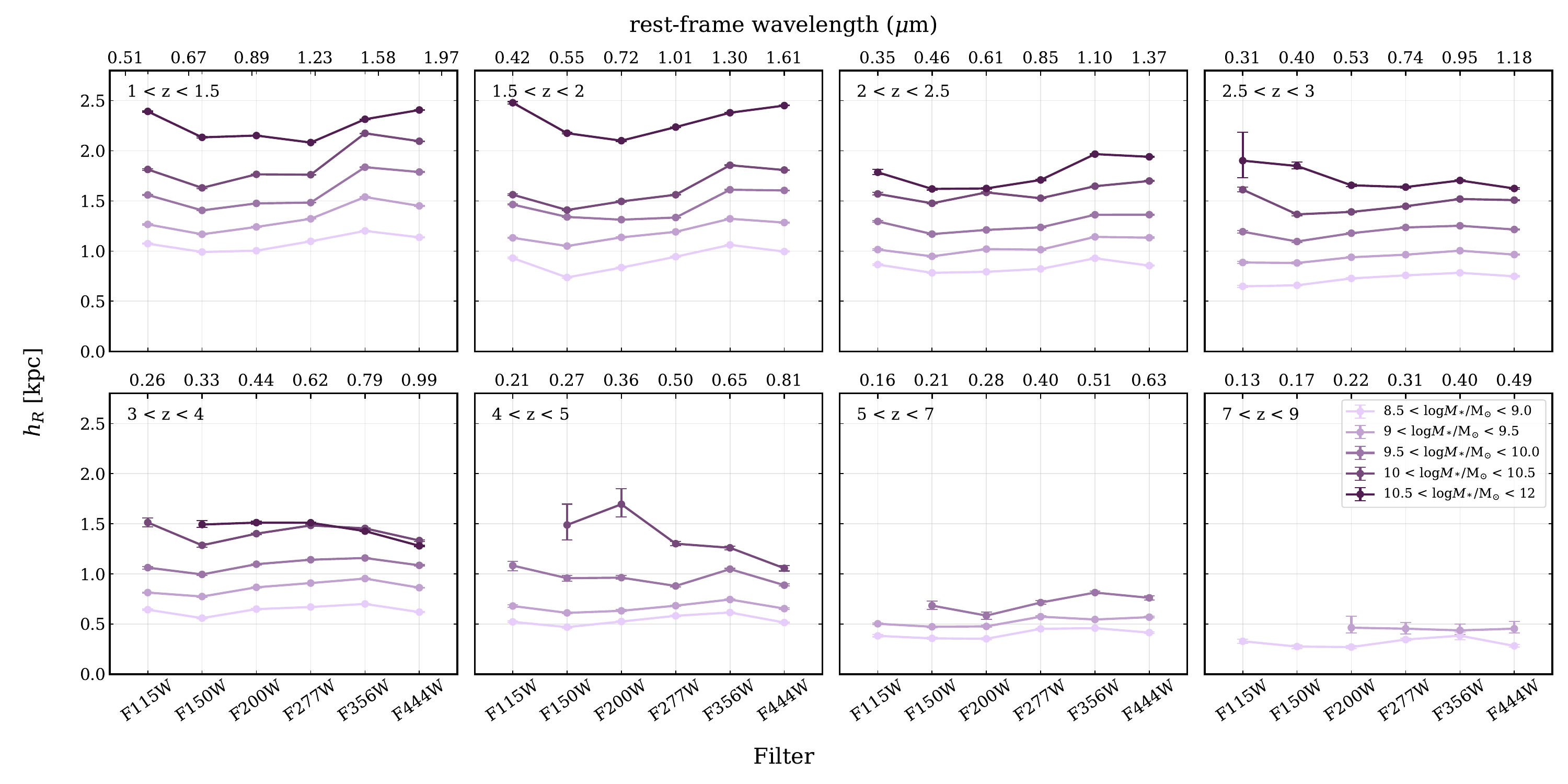}
  \caption{Outer scale lengths measured beyond $2R_{\rm e}$ for galaxies in different stellar-mass and redshift bins as a function of band wavelength. The uncertainties in $h_{\rm R}$ are estimated from 100 bootstrap realizations of the stacked images, with galaxies randomly resampled with replacement.
  }
  \label{Fig.hR}
\end{figure*}


We measure the outer scale length of each stacked galaxy by fitting the surface brightness profile beyond $2R_{\rm e}$, where $R_{\rm e}$ is measured based on the stacked image. Because the stacking process combines galaxies with a range of orientations without rotating them, the resulting stacked image is effectively azimuthally averaged, and its $R_{\rm e}$ is therefore expected to be smaller than the typical galaxy size measured along the major axis at the same stellar mass and redshift.
As discussed in Section \ref{subsec.prof}, the profiles of the most massive galaxies at $z<2$ show a noticeable flattening in the outermost regions, which may be associated with residual background effects. To avoid this feature biasing the scale-length measurements, we restrict the fitting range to $2R_{\rm e}<r<16$ kpc. The adopted fitting ranges are indicated in Figure \ref{Fig.profile}.

Figure \ref{Fig.hR} shows the outer scale lengths measured for the stacked galaxies in different bands. Because we impose an isotropic model in the {\sc Galfit} fitting, the F150W stacks of galaxies with $9<\log M_*/{\rm M_{\odot}}<9.5$ in the highest-redshift bin cannot be reliably deconvolved and reconstructed. Their relatively low signal-to-noise ratios and small numbers of galaxies result in deviations from the assumed isotropic morphology. In addition, we exclude from the scale-length analysis any band for which the stacked profile falls below the background threshold within $2R_{\rm e}$, a situation that occurs primarily in the highest-redshift bin. These cases are omitted from Figure \ref{Fig.hR}.

Overall, the outer scale lengths show no significant wavelength dependence for the vast majority of redshift and stellar-mass bins. In the three lowest-redshift bins, however, the scale lengths show a weak tendency to first decrease and then increase toward longer wavelengths, corresponding to rest-frame wavelengths of approximately $1.5~\mu{\rm m}$. Apart from the most massive mass bin, the wavelength dependence appears broadly similar across different stellar-mass bins. At $z\gtrsim2.5$, this weak increase with wavelength largely disappears. For galaxies with $M_*>10^{10}\ M_\odot$, we tentatively find a weak decrease in scale length toward longer wavelengths ($\sim 1 \ \mu{\rm m}$), most noticeably in the $4<z<5$ bins. 

The outer scale length increases with stellar mass and decreases with redshift, qualitatively resembling the evolution of galaxy sizes \citep[e.g.][]{van_der_Wel_2014,Jia_2024}. This raises the question of whether the scale length measured in the outskirts is governed by physical processes similar to those that determine the conventional half-light radius, which is dominated by the inner regions. To investigate whether the outer light represents a natural extension of the inner galaxy or instead reflects a physically distinct stellar component, we examine the ``$M_*-h_{\rm R,out}$'' relation and its evolution with redshift. Comparing this relation with the well-established mass--size relation provides a useful way to characterize the connection between the inner and outer light distributions and to gain insight into the physical origin of the faint outskirts.

\begin{figure*}[htbp]
  \centering
  \includegraphics[width=0.48\textwidth]{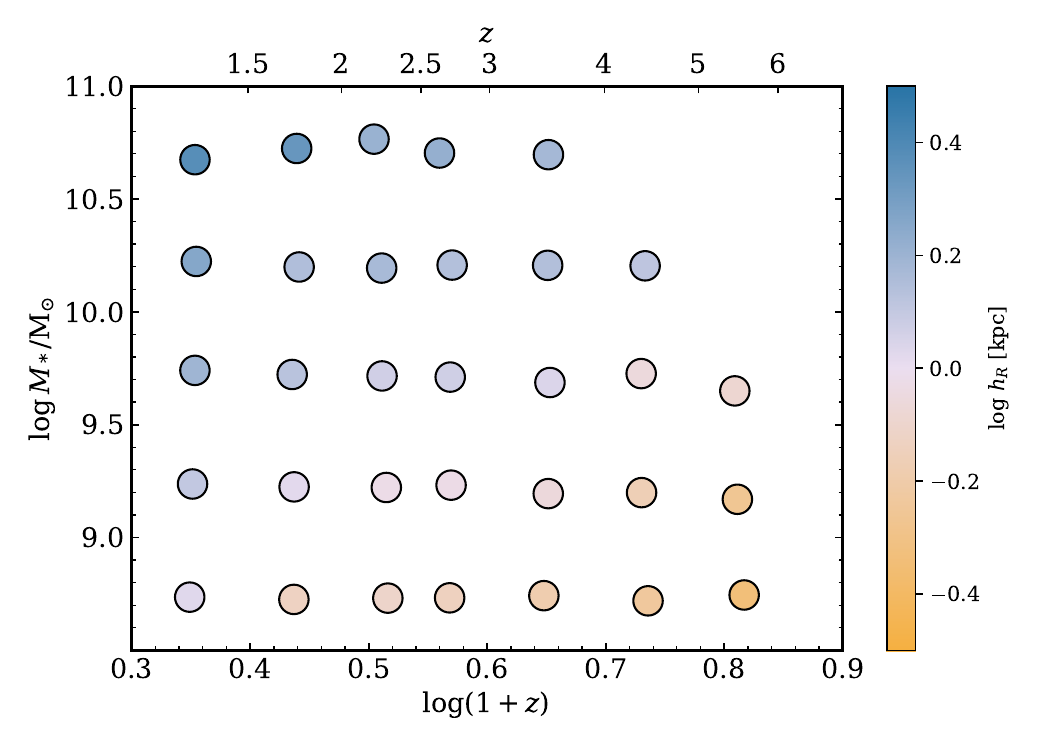}
  \includegraphics[width=0.48\textwidth]{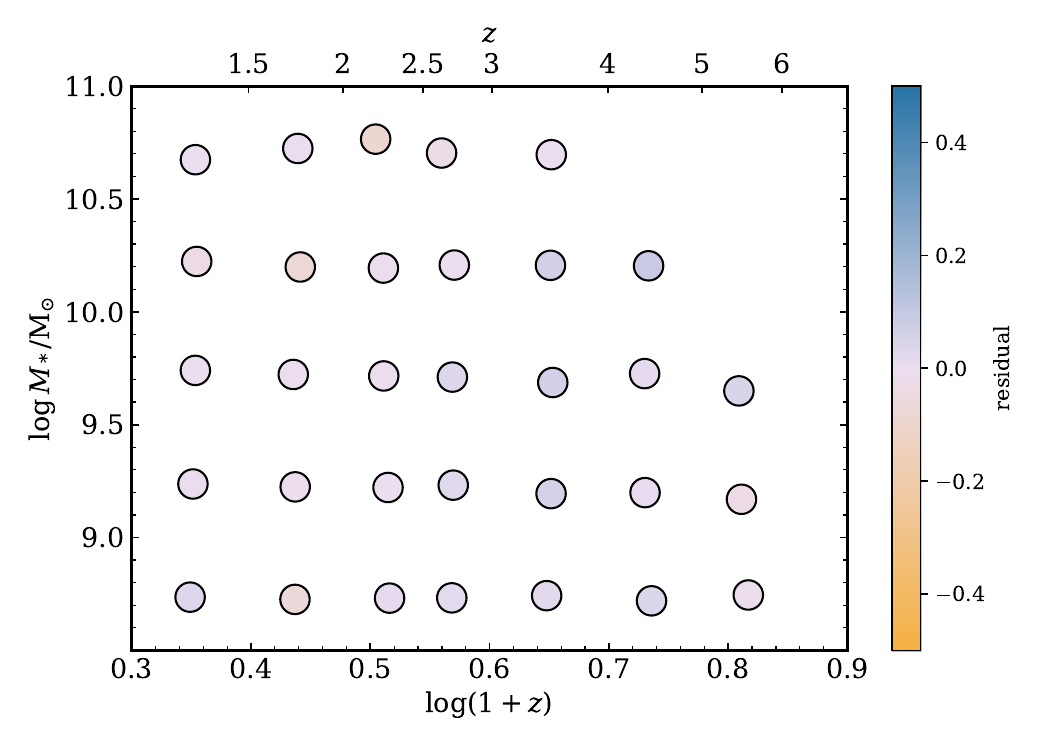}
  \caption{ The dependence of the outer scale length, $h_{\rm R}$, on stellar mass and redshift. The mass--$h_{\rm R}$ relation evolves with redshift, exhibiting slopes and trends similar to those of the conventional mass--size relation. To ensure a consistent comparison across redshifts, we select, for each redshift bin, the observed band whose effective rest-frame wavelength is closest to $0.5~\mu{\rm m}$. The right panel shows the residuals from the best-fit relation.}
  \label{Fig.hR_evo}
\end{figure*}

Figure \ref{Fig.hR_evo} shows the outer scale length, $h_{\rm R}$, as a function of stellar mass and redshift, where the stellar mass and redshift are represented by the median values of the galaxies in each bin. Although the outer scale length exhibits only a weak dependence on wavelength, as discussed above, we select, at each redshift, the observed band whose effective rest-frame wavelength is closest to $0.5~\mu{\rm m}$. This choice provides a relatively consistent tracer of the underlying stellar distribution across the full redshift range.

We fit the dependence of $h_{\rm R}$ on stellar mass and redshift using the following functional form \citep{Jia_2024}:
\begin{equation}
    \log h_{\rm R} = \alpha \log(M_*/{\rm M_{\odot}}) + \beta \log(1+z) + k,
\end{equation}
where $\alpha$ characterizes the dependence of the outer scale length on stellar mass, analogous to the slope of the conventional mass--size relation, while $\beta$ describes its redshift evolution, analogous to the exponent in the parameterization $R_{\rm e}\propto(1+z)^{\beta}$. Because the $z>7$ bins generally lack reliable measurements beyond $2R_{\rm e}$ in the bands corresponding to rest-frame wavelengths near $1~\mu{\rm m}$, we exclude these bins from the fit.

The best-fit parameters are $\alpha=0.1994_{-0.0128}^{+0.0128}$, $\beta=-0.6897_{-0.052}^{+0.0526}$, and $k=-1.5044_{-0.1333}^{+0.1352}$. The mass dependence is remarkably consistent with that of the conventional mass--size relation reported in previous studies \citep{Mowla_2019, Nedkova_2021, Ward_2023, Jia_2024, Song_2026}. The redshift evolution is also broadly consistent with previous measurements, although our best-fit $\beta$ is somewhat shallower than the evolution typically found for galaxy sizes, while remaining within the range reported in the literature.
Furthermore, we also find that the scale length of galaxies exhibits a strong correlation with their corresponding halo size. A detailed discussion of this result is provided in Appendix \ref{app.halo}.

The similarity between the $M_*-h_{\rm R}$ relation and the conventional mass--size relation may have several interpretations. One possibility is that the outer light at these radii is physically connected to the same stellar component and growth processes that shape the inner galaxy, rather than being dominated by ex-situ accretion. In particular, if these star-forming galaxies are dominated by disk components, the evolution of $R_{\rm e}$ derived from single-Sérsic fitting may naturally be consistent with the evolution of the scale length of the outer disk, suggesting that the inner and outer regions are part of the same continuously growing disk component.
One scenario to achieve such a simultaneous growth of the inner and outer stellar components, as recently proposed by \cite{Mo2024}, is star formation in dense sub-clouds formed as a consequence of the global self-gravitation of the entire galaxy.
Alternatively, the inner and outer components could have different origins but nevertheless be shaped by common post-formation processes that produce similar radial distributions \citep{Yu_2025}. Such processes may include stellar scattering driven by gravitational instabilities \citep{Bournaud_2007}, as well as radial redistribution induced by spiral arms and bars \citep{Roskar_2008, Minchev_2012}. However, it remains unclear whether these processes have sufficient time to significantly reshape the stellar distribution in the very early Universe. 
The smooth color profiles extending from the inner regions to the outskirts, together with the generally bluer colors of the outer regions (Figure \ref{Fig.color}), favor the former interpretation. We therefore find no evidence that the light beyond $2R_{\rm e}$ provides a direct tracer of the ex-situ stellar component at these redshifts. 
We note that \cite{He_2025}, based on simulation results, pointed out that the stellar halo contribution in face‑on galaxies might be small and overwhelmed by the light of the galactic disk.
Nevertheless, the similarity between the outer scale-length relation and the conventional mass--size relation suggests that caution is needed when using extended light as a proxy for the ex-situ stellar fraction in the early universe.


\section{summary}\label{sec.sum}


We investigate the faint outer light profiles of star-forming galaxies at $1<z<9$ using stacked JWST/NIRCam images, with the aim of characterizing the connection between the outer regions and the inner galaxy and assessing their potential relation to the emergence of stellar halos. Our analysis is based on deep imaging from the CEERS, JADES, and PRIMER surveys, covering observed wavelengths from $\sim1.1$ to $4.4~\mu{\rm m}$. We construct a sample of 14,812 galaxies with reliable imaging and structural measurements, selected to have minor-to-major axis ratios greater than 0.5. Using multi-band stacked images, we characterize the radial color profiles and outer light distributions, investigate their dependence on stellar mass and redshift, and explore the possible physical origin of the light at these radii.

We summarize our findings as follows:

\begin{enumerate}



    \item The stacked surface brightness profiles remain smooth and well detected to radii beyond $\sim5-10$ kpc. For the most massive galaxies at $z\sim1.5$, we detect the outer light to $\sim20$ kpc at the $3\sigma$ level. At all redshifts, more massive galaxies exhibit more extended outer light profiles, while the detectable radius decreases rapidly with increasing redshift, from $\sim20$ kpc at $z\sim1.5$ to $\sim5$ kpc at $z\sim4$. This trend is broadly consistent with the known size evolution of high-redshift galaxies.

    \item The radial color profiles generally show a smooth red-to-blue gradient from the center toward the outskirts. This trend is observed over a broad range of redshifts and stellar masses, with more massive galaxies being systematically redder, qualitatively consistent with previous findings at lower redshifts. The observed gradients are broadly consistent with an inside-out growth scenario, although variations in dust attenuation may also contribute. At $z\gtrsim2.5$, low-mass galaxies show a tentative reversal of this trend in their outermost regions, with the colors becoming slightly redder toward larger radii. This may suggest a different mode of galaxy growth, with low-mass galaxies at high redshift preferentially undergoing outside-in growth.

    \item The outer scale lengths measured beyond $2R_{\rm e}$ show a mass and redshift dependence remarkably similar to that of the conventional mass--size relation, with very weak wavelength dependence. The similarity in their scaling relations may suggest that the outer light is a natural extension of the inner stellar component rather than being dominated by a distinct ex-situ component in the early universe. However, this interpretation remains tentative, as different stellar components could also be shaped by similar physical processes. The weak wavelength dependence of the outer scale lengths further suggests that dust attenuation has only a limited effect at these radii.

\end{enumerate}


By extending measurements beyond the effective radius into the faintest regions accessible with current JWST imaging, this work provides a phenomenological characterization of the outer light profiles of high-redshift galaxies and establishes an observational bridge between their well-studied inner regions and the more extended stellar halos that develop at later times. While the current JWST data do not yet provide sufficient depth and sample size to directly probe the physical scales typically associated with stellar halos, our measurements offer a first step toward tracing the emergence and evolution of extended stellar structures from cosmic dawn and cosmic noon to the present day.

A direct characterization of stellar halos at high redshift will require deeper imaging and substantially larger samples. Future wide-field surveys, such as Euclid, together with deeper JWST observations, will provide the sensitivity, area, and statistics needed to extend these measurements to larger physical scales and to investigate the transition from the outer regions studied here to bona fide stellar halos. Such observations will ultimately enable a more direct assessment of the role of ex-situ accretion and other processes in shaping the extended stellar structures of galaxies across cosmic time.

\section{acknowledgements}\label{sec:acknowledgements}
EW thanks support of the National Science Foundation of China(Nos. 12633004 and 12473008) and the Start-up Fund of the University of Science and Technology of China(No. KY2030000200). The authors gratefully acknowledge the support of Cyrus Chun Ying Tang Foundations.
CL thanks support of the Fundamental Research Funds for the Central Universities (WK2030250123) and China Postdoctoral Science Foundation (2026T190848).

\appendix

\section{stacked images}\label{app.imgs}

\begin{figure*}[ht]
    \centering
    \includegraphics[scale=0.33]{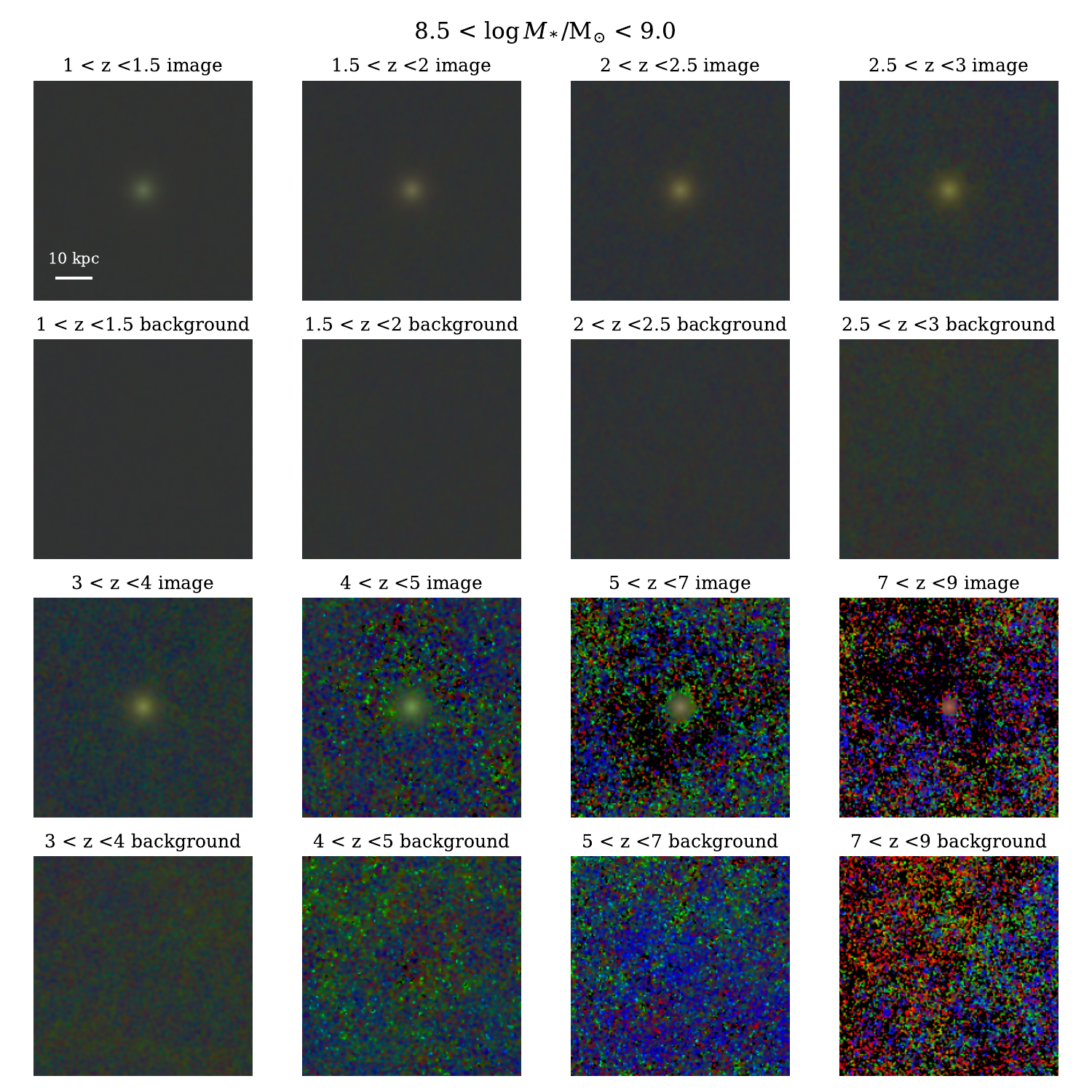}
    \includegraphics[scale=0.33]{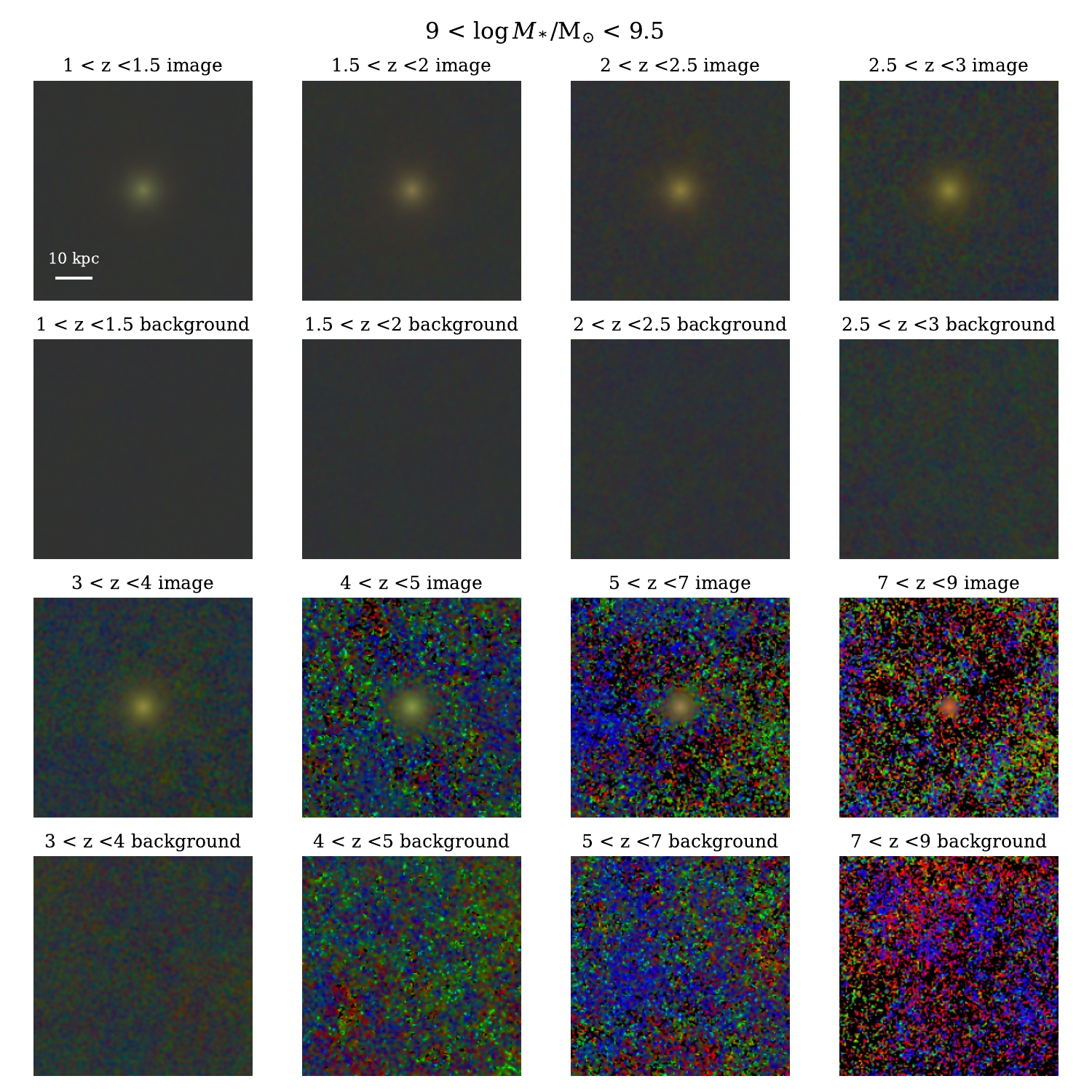}
    \caption{Same as Figure \ref{Fig.examples}, but for $8.5 < \log M*/{\rm M_{\odot}} < 9$ (left panel) and $9 < \log M*/{\rm M_{\odot}} < 9.5$ (right panel).}
    \label{Fig.examples_app1}
\end{figure*}

\begin{figure*}[ht]
    \centering
    \includegraphics[scale=0.33]{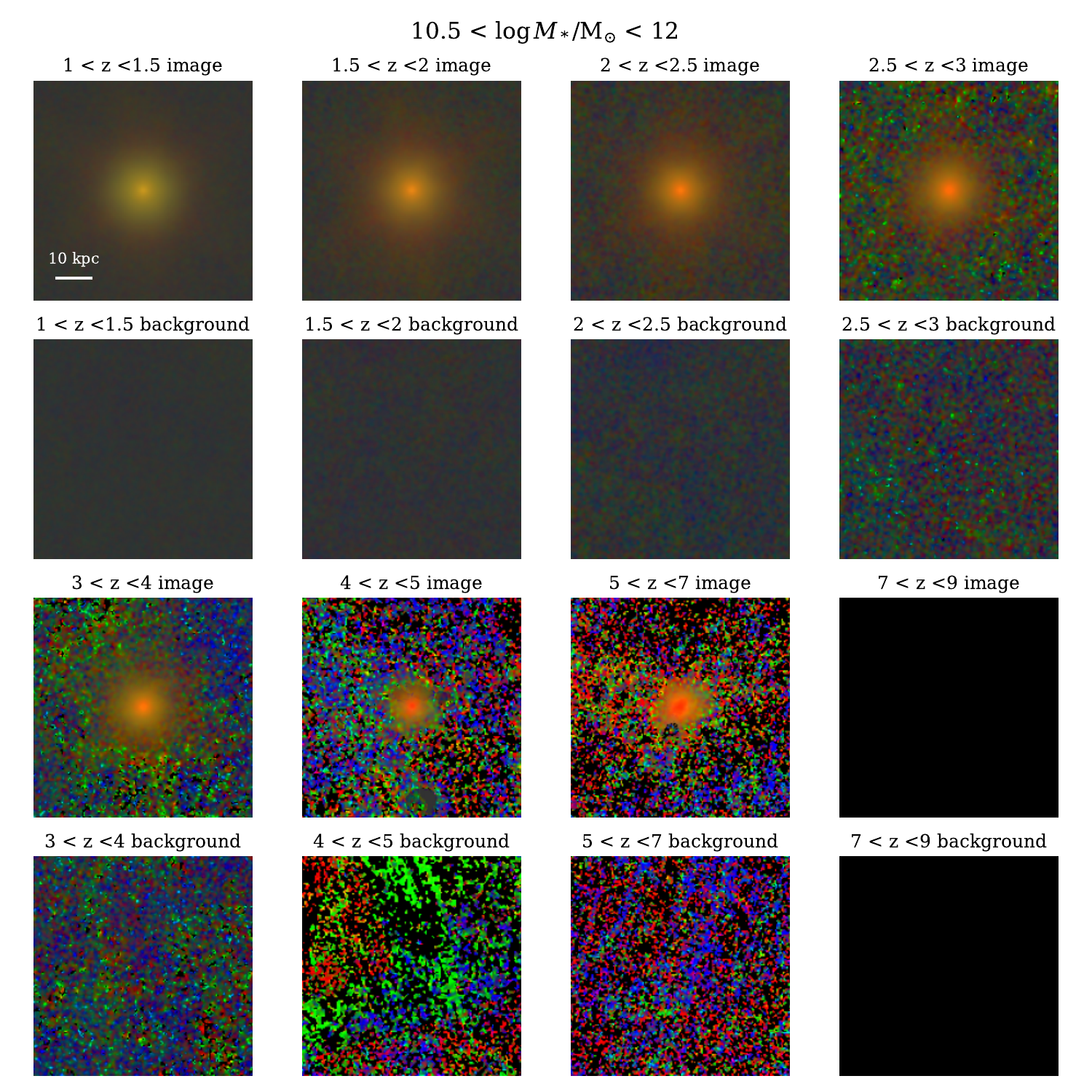}
    \caption{Same as Figure \ref{Fig.examples}, but for $10.5 < \log M*/{\rm M_{\odot}} < 12$.}
    \label{Fig.examples_app2}
\end{figure*}

\section{surface brightness profiles}\label{app.prof}

Here we present the surface brightness profiles in the F115W, F150W, F200W, F277W, and F356W bands, which are analogous to the F444W profiles shown in Figure \ref{Fig.profile} in the main text. The corresponding data tables are available at Data behind the Figure.

\begin{figure*}[htbp]
  \centering
  \includegraphics[width=0.95\textwidth]{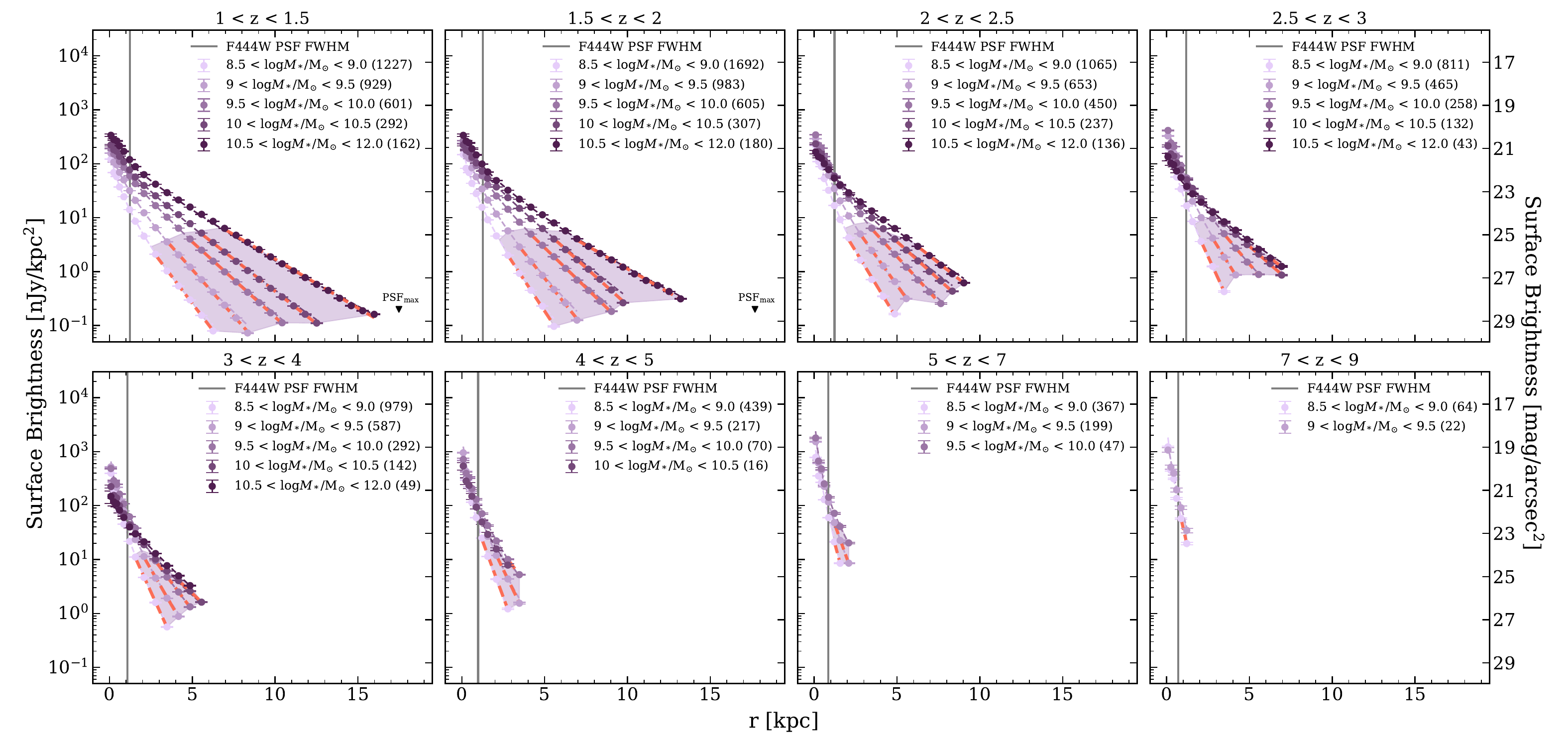}
  \caption{Same as Figure \ref{Fig.profile}, but for the F115W band.}
  \label{Fig.profile_115}
\end{figure*}

\begin{figure*}[htbp]
  \centering
  \includegraphics[width=0.95\textwidth]{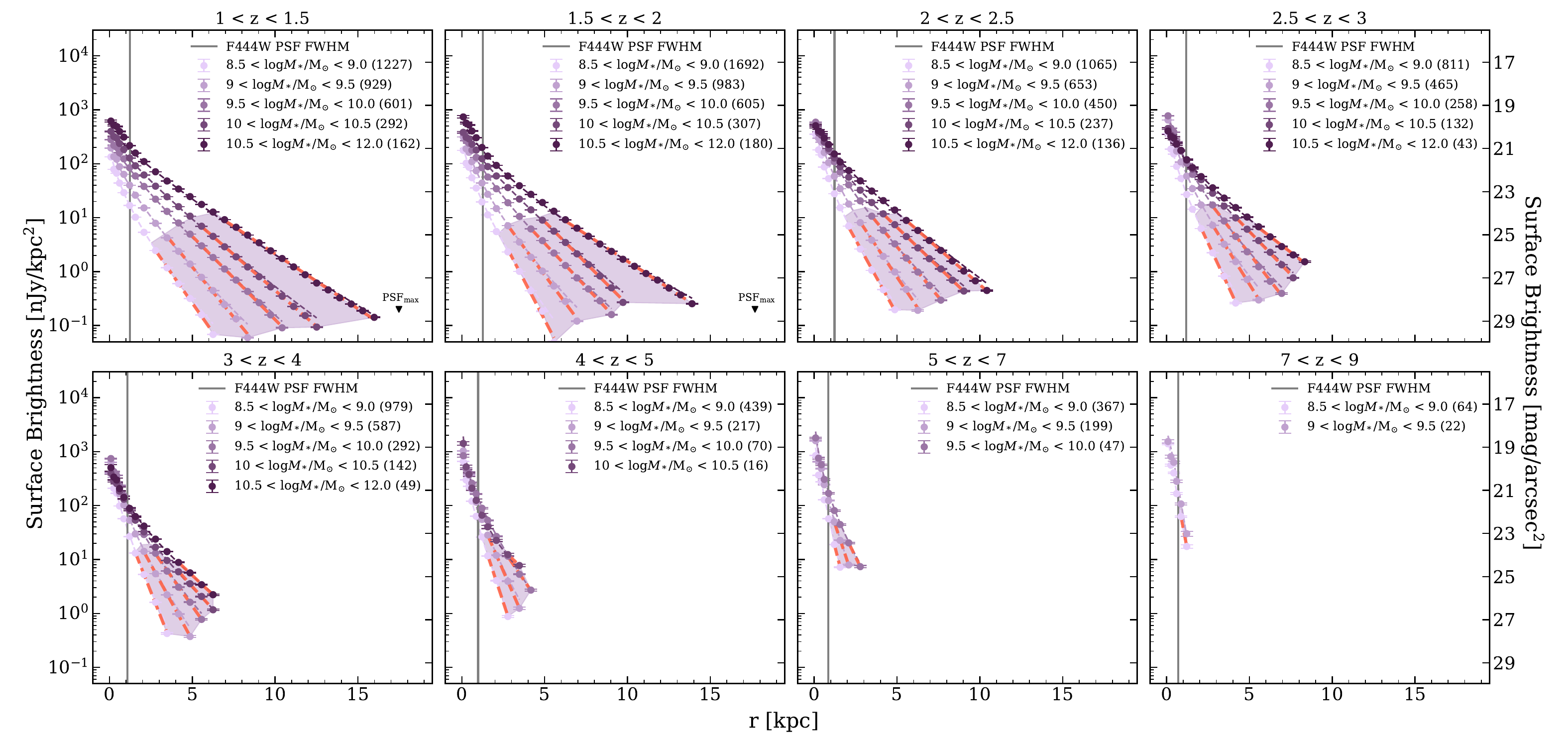}
  \caption{Same as Figure \ref{Fig.profile}, but for the F150W band.}
  \label{Fig.profile_150}
\end{figure*}

\begin{figure*}[htbp]
  \centering
  \includegraphics[width=0.95\textwidth]{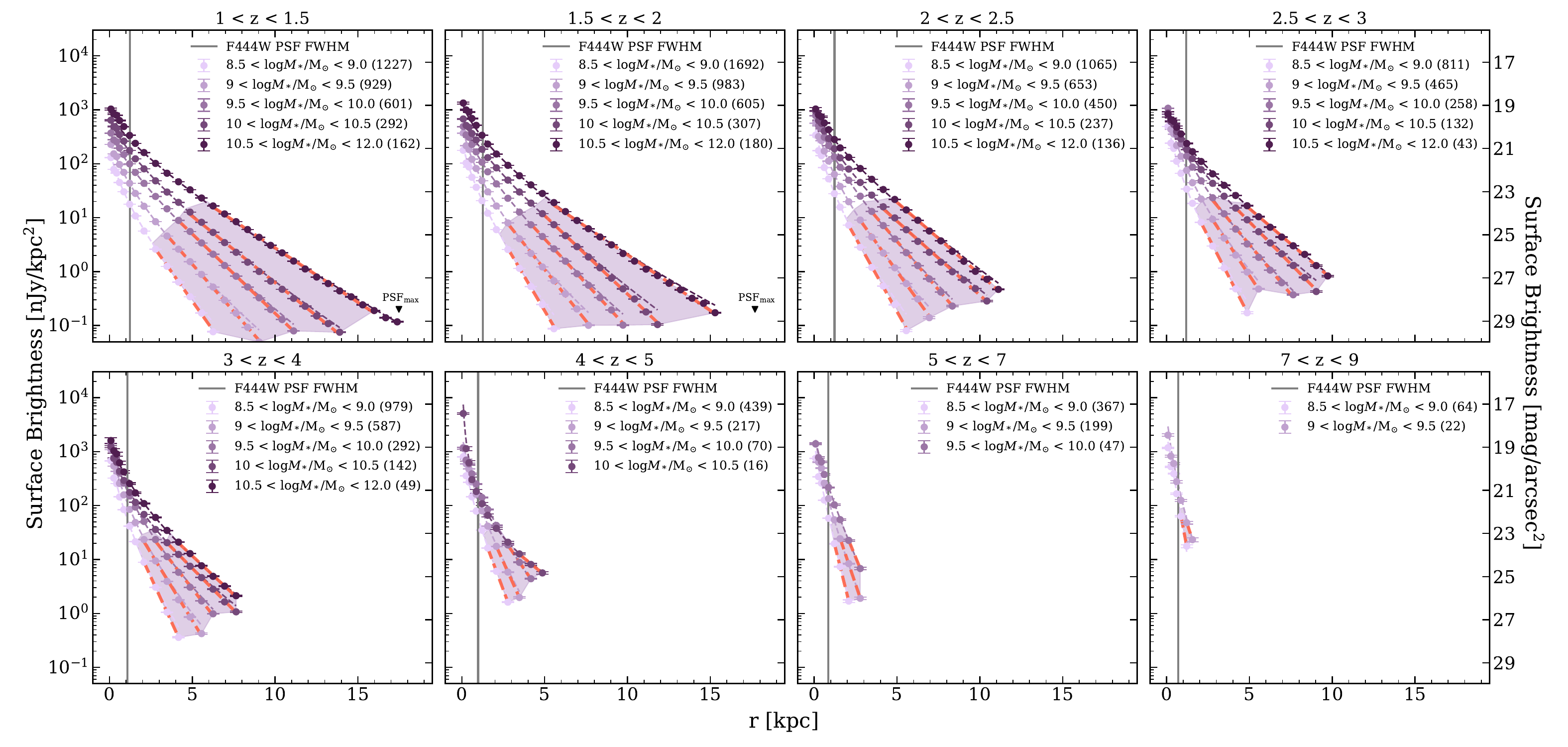}
  \caption{Same as Figure \ref{Fig.profile}, but for the F200W band.}
  \label{Fig.profile_200}
\end{figure*}

\begin{figure*}[htbp]
  \centering
  \includegraphics[width=0.95\textwidth]{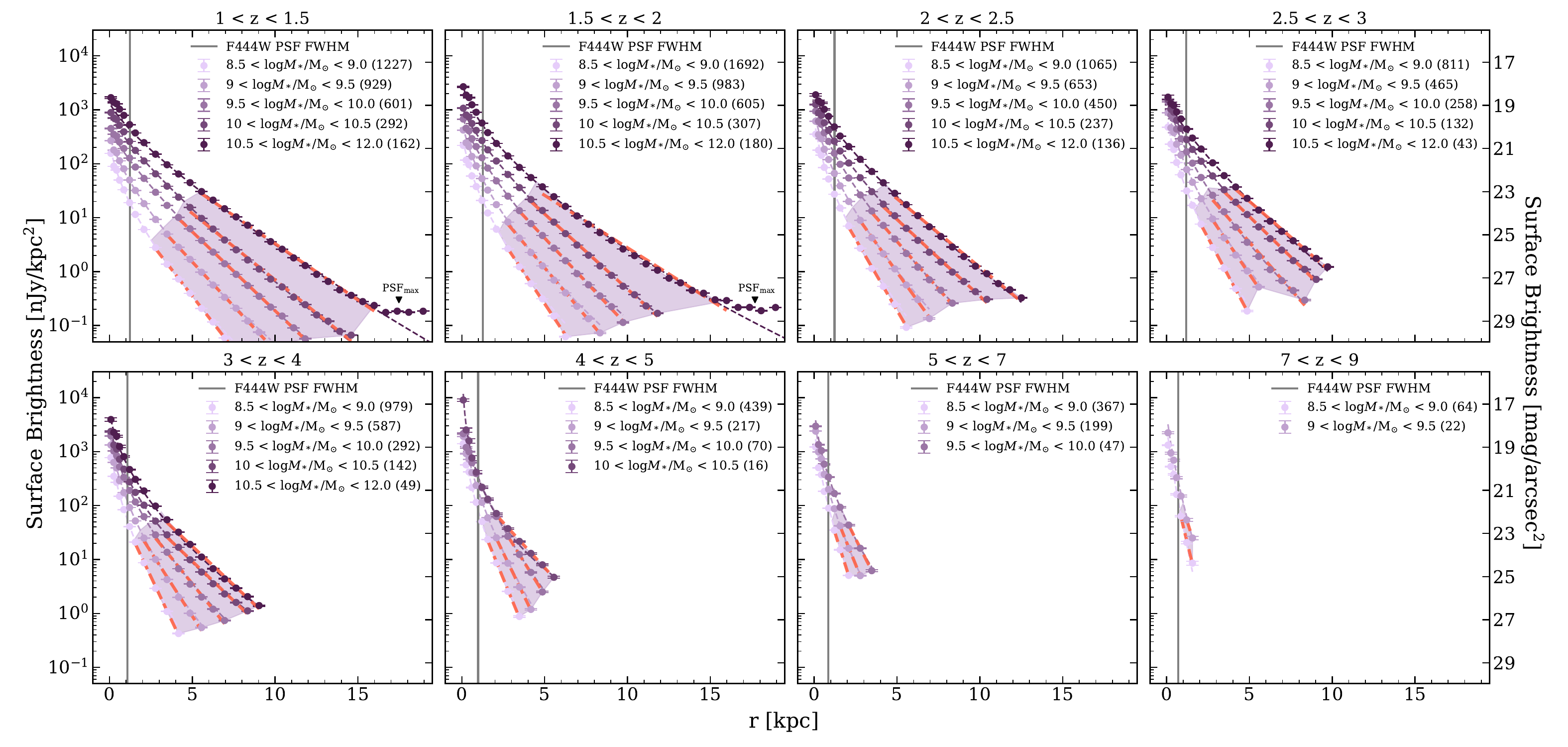}
  \caption{Same as Figure \ref{Fig.profile}, but for the F277W band.}
  \label{Fig.profile_277}
\end{figure*}

\begin{figure*}[htbp]
  \centering
  \includegraphics[width=0.95\textwidth]{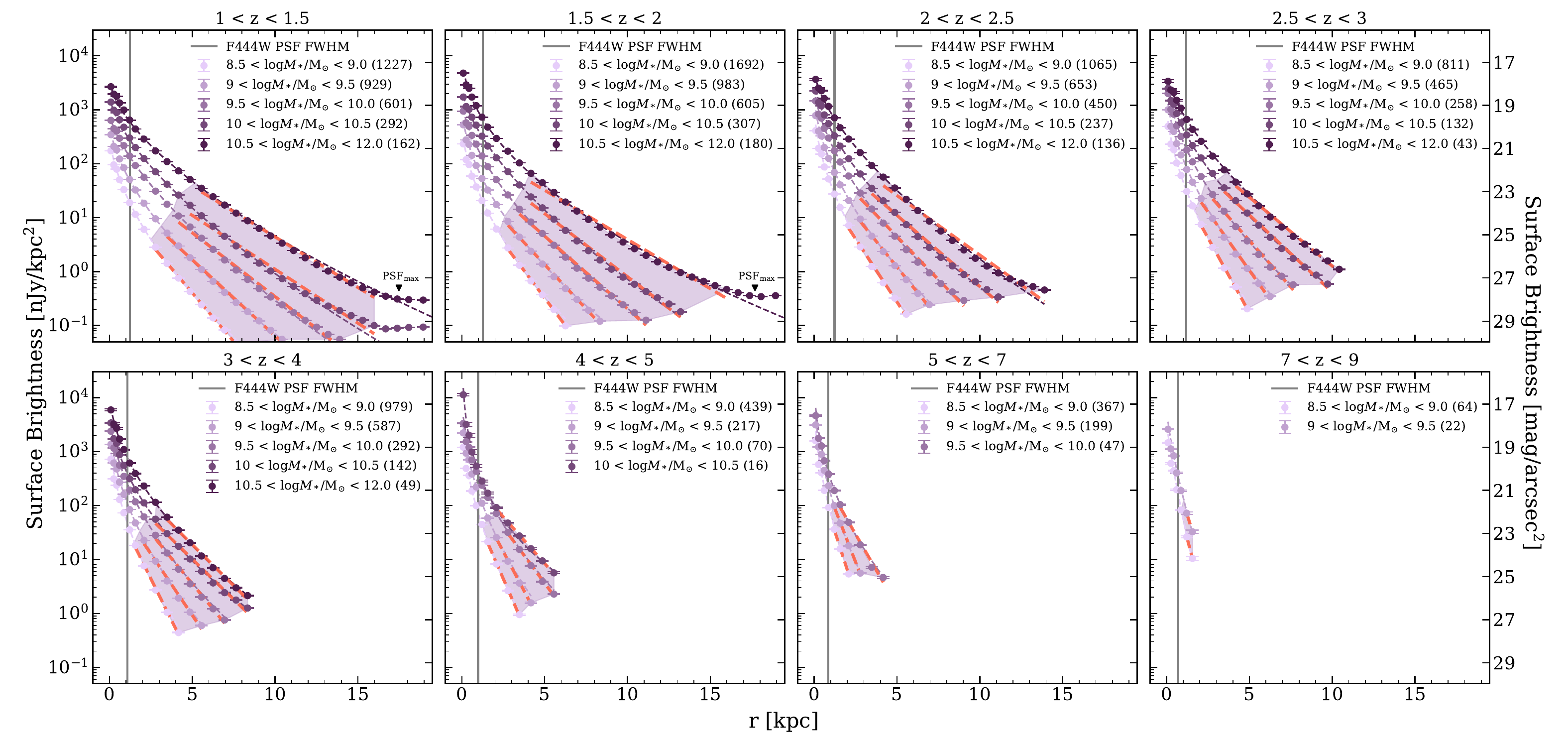}
  \caption{Same as Figure \ref{Fig.profile}, but for the F356W band.}
  \label{Fig.profile_356}
\end{figure*}

\section{The impact of missing CEERS}\label{app.ceers}


To match the rest-frame wavelengths as closely as possible, we use the F090W photometry when constructing the color indices. However, F090W imaging is not available in the CEERS field, and CEERS sources are therefore excluded from the F090W stacks. For the other bands, we include all available sources from CEERS, JADES, and PRIMER to maximize the signal-to-noise ratio. To assess whether the absence of CEERS data introduces any systematic bias, we compare the surface brightness profiles constructed with and without CEERS sources in the two lowest-redshift bins, $1<z<1.5$ and $1.5<z<2$, where F090W is used for the color measurements. We find no significant systematic difference between the two cases, indicating that the exclusion of CEERS does not bias our color measurements. This further supports the robustness of our results and suggests that they are not significantly affected by cosmic variance. 

\begin{figure*}[htbp]
  \centering
  \includegraphics[width=0.95\textwidth]{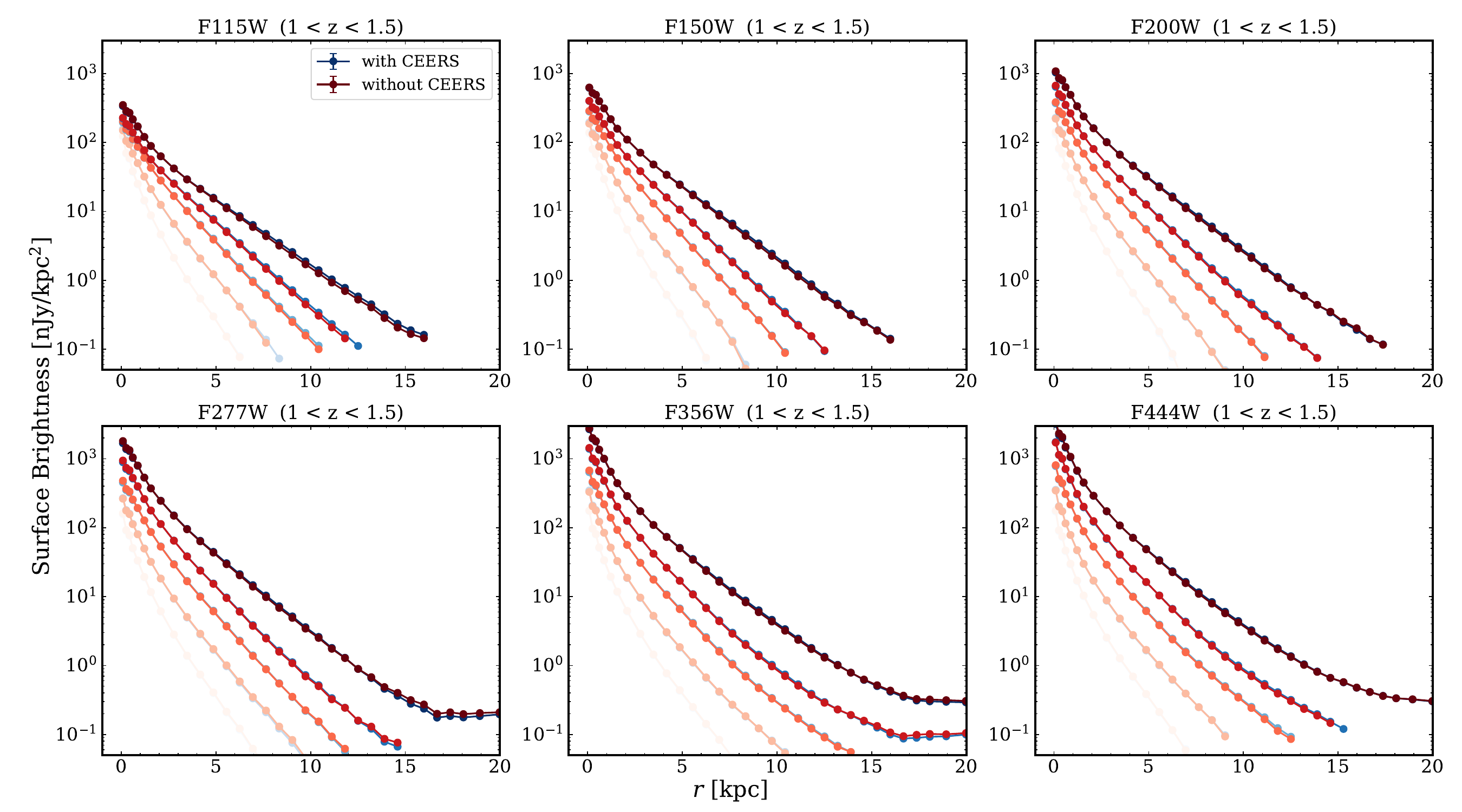}
  \includegraphics[width=0.95\textwidth]{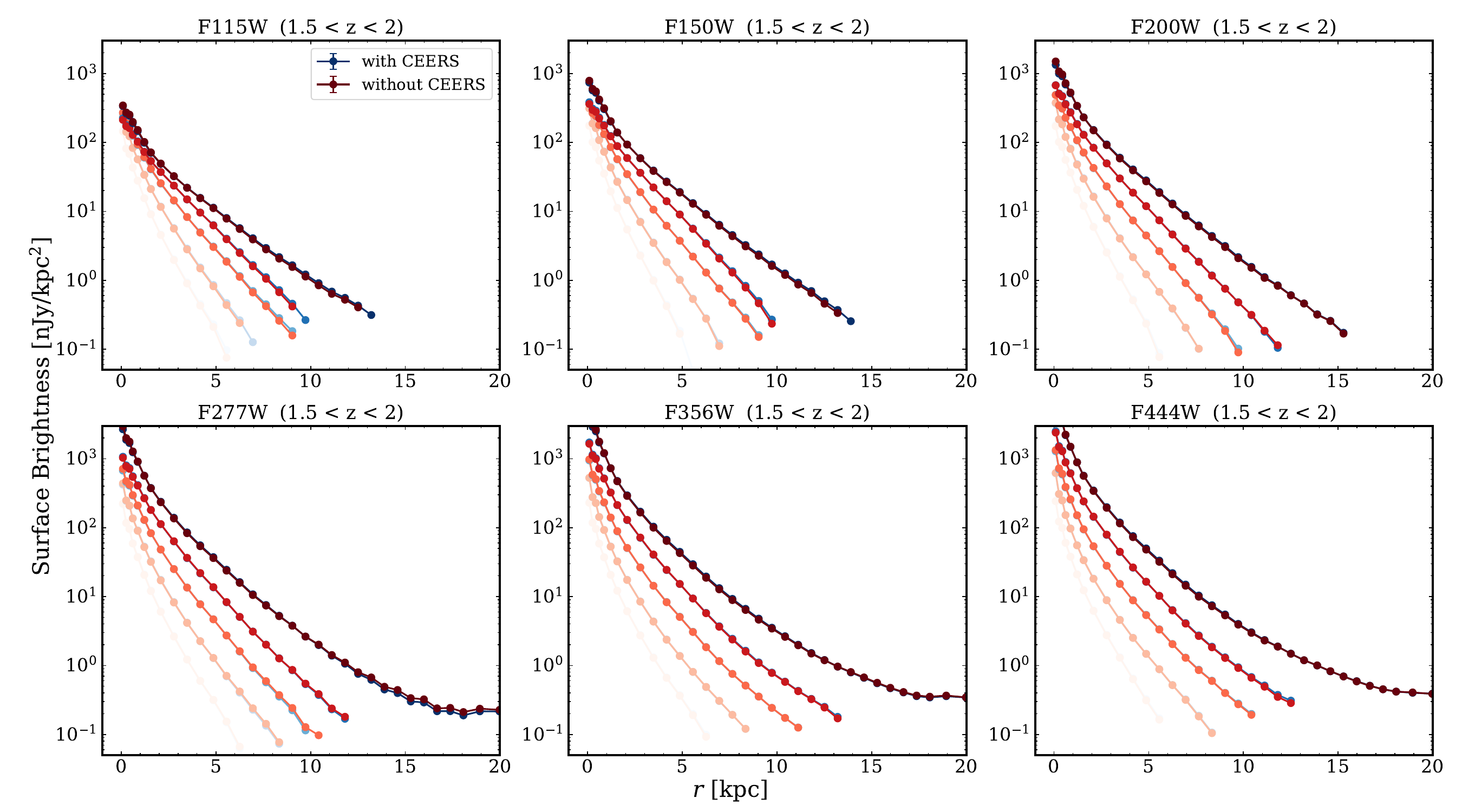}
  \caption{Comparison of the surface brightness profiles measured with and without the CEERS field for different bands illustrated in the two redshift bins, $1.0<z<1.5$ and $1.5<z<2$. The profiles are nearly identical in both cases, indicating that excluding the CEERS field does not significantly affect the results. }
  \label{Fig.noceers}
\end{figure*}

\section{dark matter halo size}\label{app.halo}

\begin{figure*}[htbp]
  \centering
  \includegraphics[width=0.95\textwidth]{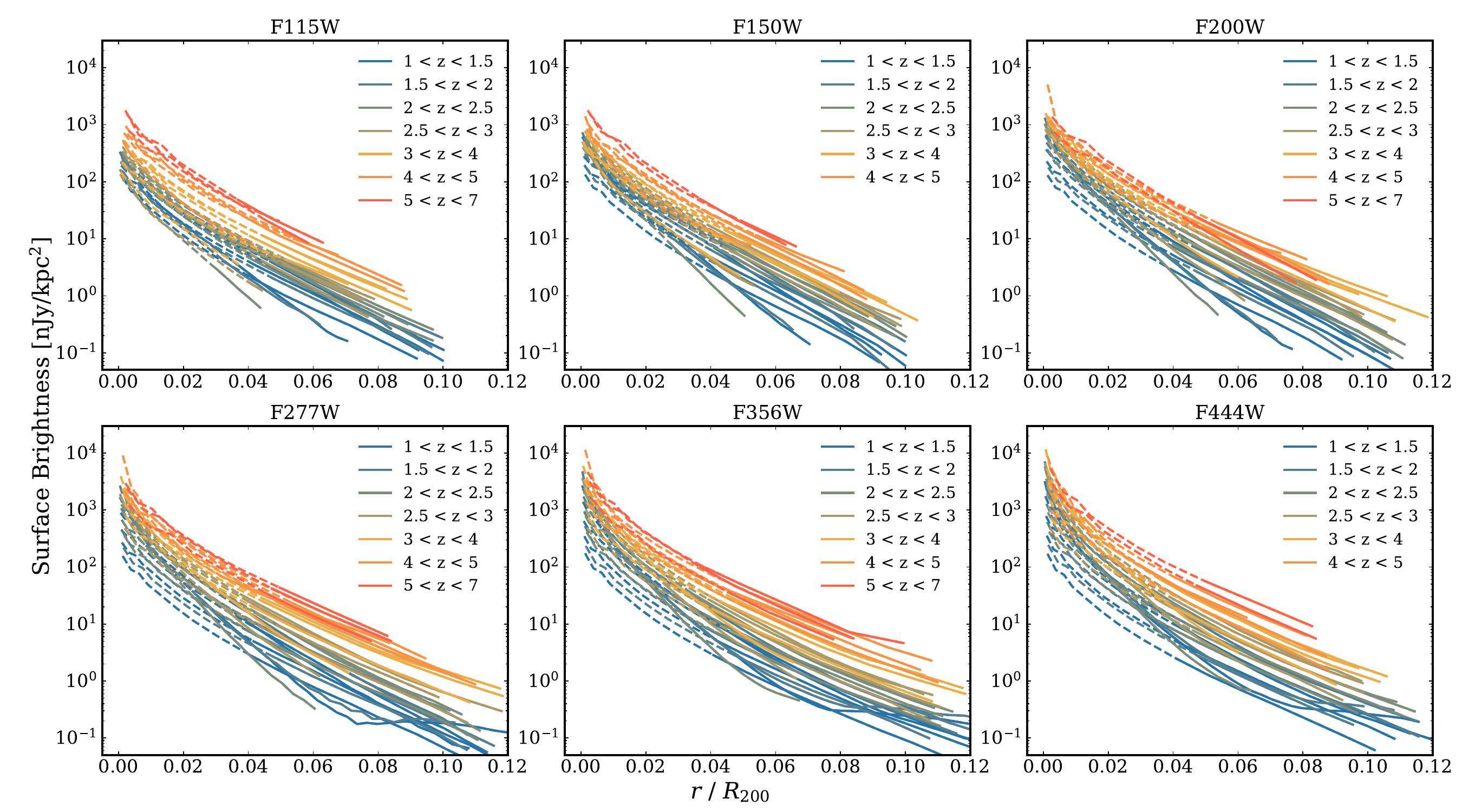}
  \caption{Normalized galaxy surface brightness profiles as a function of radius in units of $R_{200}$ for different bands and redshift bins. Each panel shows a different band, with profiles at different redshifts distinguished by color. For clarity, we do not distinguish between stellar mass bins using colors or line styles, as more massive galaxies are systematically brighter at a given redshift. The inner $2R_{\rm e}$ region is indicated by dashed lines. }
  \label{Fig.halo}
\end{figure*}

An interesting finding of our analysis, consistent with recent stacking studies of galaxy outskirts \citep{WangWT_2019}, is that the outer stellar light profiles exhibit a remarkable degree of self-similarity when the projected radius is normalized by the halo virial radius , $R_{200}$. Here, we estimate $R_{200}$ using the stellar mass--halo mass relation from \cite{Moster_2013}. As shown in Figure \ref{Fig.halo}, when the radial coordinate is expressed in units of $R_{200}$ rather than physical kiloparsecs, galaxies spanning a broad range of stellar masses ($8.5 < \log M_*/{\rm M_{\odot}} < 12$) exhibit remarkably similar outer surface brightness profiles, with comparable slopes over a radial range of $\sim0.1R_{200}$. This result is consistent with the findings of \cite{WangWT_2019} and appears to hold over a broad redshift range.

The observed self-similarity may indicate that the stellar distribution on these scales is closely related to the assembly of the host dark matter halo. Given the potentially complex origins of stars in the outskirts, it may also point to an important contribution from externally acquired stellar material. Alternatively, a common post-formation process could shape the radial distribution of stars in a manner that scales with halo properties. If confirmed, the persistence of this self-similarity across a wide range of stellar masses and redshifts would require a physical mechanism capable of producing a remarkably uniform radial structure, such as dynamical heating associated with dark matter halo assembly \citep{Chen_2024}.

We emphasize, however, that the stellar mass--halo mass relation at high redshift remains poorly constrained observationally. Moreover, our measurements probe only a limited radial range and do not directly establish the origin of the stars. We therefore regard this self-similarity as an intriguing observational trend rather than evidence for a specific physical mechanism, and leave a more detailed interpretation to future studies.

\bibliography{ref}

@misc{ceers_2025,
      title={The Cosmic Evolution Early Release Science Survey (CEERS)}, 
      author={Steven L. Finkelstein and Micaela B. Bagley and Pablo Arrabal Haro and Mark Dickinson and Henry C. Ferguson and Jeyhan S. Kartaltepe and Dale D. Kocevski and Anton M. Koekemoer and Jennifer M. Lotz and Casey Papovich and Pablo G. Perez-Gonzalez and Nor Pirzkal and Rachel S. Somerville and Jonathan R. Trump and Guang Yang and L. Y. Aaron Yung and Adriano Fontana and Andrea Grazian and Norman A. Grogin and Lisa J. Kewley and Allison Kirkpatrick and Rebecca L. Larson and Laura Pentericci and Swara Ravindranath and Stephen M. Wilkins and Omar Almaini and Ricardo O. Amorin and Guillermo Barro and Rachana Bhatawdekar and Laura Bisigello and Madisyn Brooks and Fernando Buitrago and Antonello Calabro and Marco Castellano and Yingjie Cheng and Nikko J. Cleri and Justin W. Cole and M. C. Cooper and Olivia R. Cooper and Luca Costantin and Isa G. Cox and Darren Croton and Emanuele Daddi and Kelcey Davis and Avishai Dekel and David Elbaz and Vital Fernandez and Seiji Fujimoto and Giovanni Gandolfi and Jonathan P. Gardner and Eric Gawiser and Mauro Giavalisco and Carlos Gomez-Guijarro and Yuchen Guo and Ansh R. Gupta and Nimish P. Hathi and Santosh Harish and Aurelien Henry and Michaela Hirschmann and Weida Hu and Taylor A. Hutchison and Kartheik G. Iyer and Anne E. Jaskot and Saurabh W. Jha and Intae Jung and Vasily Kokorev and Peter Kurczynski and Gene C. K. Leung and Mario Llerena and Arianna S. Long and Ray A. Lucas and Shiying Lu and Elizabeth J. McGrath and Daniel H. McIntosh and Emiliano Merlin and Alexa M. Morales and Lorenzo Napolitano and Fabio Pacucci and Viraj Pandya and Marc Rafelski and Giulia Rodighiero and Caitlin Rose and Paola Santini and Lise-Marie Seille and Raymond C. Simons and Lu Shen and Amber N. Straughn and Sandro Tacchella and Brittany N. Vanderhoof and Jesus Vega-Ferrero and Benjamin J. Weiner and Christopher N. A. Willmer and Peixin Zhu and Eric F. Bell and Stijn Wuyts and Benne W. Holwerda and Xin Wang and Weichen Wang and Jorge A. Zavala},
      year={2025},
      eprint={2501.04085},
      archivePrefix={arXiv},
      primaryClass={astro-ph.GA},
      url={https://arxiv.org/abs/2501.04085}, 
}

@ARTICLE{Wang2018,
       author = {{Wang}, Enci and {Li}, Cheng and {Xiao}, Ting and {Lin}, Lin and {Bershady}, Matthew and {Law}, David R. and {Merrifield}, Michael and {Sanchez}, Sebastian F. and {Riffel}, Rogemar A. and {Riffel}, Rogerio and {Yan}, Renbin},
        title = "{SDSS-IV MaNGA: Star Formation Cessation in Low-redshift Galaxies. I. Dependence on Stellar Mass and Structural Properties}",
      journal = {\apj},
         year = 2018,
        month = apr,
       volume = {856},
       number = {2},
          eid = {137},
        pages = {137},
          doi = {10.3847/1538-4357/aab263},
archivePrefix = {arXiv},
       eprint = {1710.07569},
 primaryClass = {astro-ph.GA},
       adsurl = {https://ui.adsabs.harvard.edu/abs/2018ApJ...856..137W}
}

@ARTICLE{Wang2020,
       author = {{Wang}, Enci and {Lilly}, Simon J.},
        title = "{The Variability of the Star Formation Rate in Galaxies. I. Star Formation Histories Traced by EW(H{\ensuremath{\alpha}}) and EW(H{\ensuremath{\delta}}$_{A}$)}",
      journal = {\apj},
         year = 2020,
        month = apr,
       volume = {892},
       number = {2},
          eid = {87},
        pages = {87},
          doi = {10.3847/1538-4357/ab7b7d},
archivePrefix = {arXiv},
       eprint = {1912.06523},
 primaryClass = {astro-ph.GA},
       adsurl = {https://ui.adsabs.harvard.edu/abs/2020ApJ...892...87W}
}

@ARTICLE{LiHao2025,
       author = {{Li}, Hao and {Chen}, Yangyao and {Wang}, Huiyuan and {Mo}, Houjun},
        title = "{Physical processes behind the co-evolution of haloes, galaxies, and supermassive black holes in the IllustrisTNG simulation}",
      journal = {\mnras},
         year = 2025,
        month = oct,
       volume = {543},
       number = {2},
        pages = {1878-1898},
          doi = {10.1093/mnras/staf1594},
archivePrefix = {arXiv},
       eprint = {2409.06208},
 primaryClass = {astro-ph.GA},
       adsurl = {https://ui.adsabs.harvard.edu/abs/2025MNRAS.543.1878L}
}

@ARTICLE{Ma2026,
       author = {{Ma}, Qinglin and {Chen}, Yangyao and {Mo}, Houjun},
        title = "{Two-phase formation of galaxies: the coevolution between galaxies and dark matter haloes}",
      journal = {\mnras},
         year = 2026,
        month = may,
       volume = {548},
       number = {3},
          eid = {stag562},
        pages = {stag562},
          doi = {10.1093/mnras/stag562},
archivePrefix = {arXiv},
       eprint = {2509.16138},
 primaryClass = {astro-ph.GA},
       adsurl = {https://ui.adsabs.harvard.edu/abs/2026MNRAS.548ag562M}
}

@ARTICLE{Chen2025,
       author = {{Chen}, Yangyao and {Mo}, Houjun and {Wang}, Huiyuan},
        title = "{A two-phase model of galaxy formation: IV. Seeding and growing supermassive black holes in dark matter halos}",
      journal = {arXiv e-prints},
         year = 2025,
        month = sep,
          eid = {arXiv:2509.03283},
        pages = {arXiv:2509.03283},
          doi = {10.48550/arXiv.2509.03283},
archivePrefix = {arXiv},
       eprint = {2509.03283},
 primaryClass = {astro-ph.GA},
       adsurl = {https://ui.adsabs.harvard.edu/abs/2025arXiv250903283C}
}

@ARTICLE{Miller2021,
       author = {{Miller}, Tim B. and {van Dokkum}, Pieter},
        title = "{Bayesian Fitting of Multi-Gaussian Expansion Models to Galaxy Images}",
      journal = {\apj},
         year = 2021,
        month = dec,
       volume = {923},
       number = {1},
          eid = {124},
        pages = {124},
          doi = {10.3847/1538-4357/ac2b30},
archivePrefix = {arXiv},
       eprint = {2109.13262},
 primaryClass = {astro-ph.GA},
       adsurl = {https://ui.adsabs.harvard.edu/abs/2021ApJ...923..124M}
}

@ARTICLE{Ma2026b,
       author = {{Ma}, Chengyu and {Chen}, Yangyao and {Wang}, Enci and {Mo}, Houjun and {Wang}, Huiyuan and {Wang}, Tao and {Yuan}, Qirong and {Du}, Min and {Li}, Zhaozhou and {Wang}, Kai and {Lyu}, Cheqiu and {Yu}, Haoran and {Chen}, Zeyu and {Chen}, Xiaoxuan},
        title = "{Fountain pattern of baryon cycle revealed in galaxy ecosystems}",
      journal = {arXiv e-prints},
         year = 2026,
        month = sep,
          eid = {arXiv:2609.21374},
        pages = {arXiv:2609.21374},
          doi = {10.48550/arXiv.2609.21374},
archivePrefix = {arXiv},
       eprint = {2609.21374},
 primaryClass = {astro-ph.GA},
       adsurl = {https://ui.adsabs.harvard.edu/abs/2026arXiv260921374M}
}

@ARTICLE{Lyu2026,
       author = {{Lyu}, Cheqiu and {Wang}, Enci and {Chen}, Zeyu and {Ma}, Chengyu and {Chen}, Yangyao and {Yu}, Haoran and {Kong}, Xu},
        title = "{Observational Evidence for Anisotropic Metal Excess around Galaxies}",
      journal = {arXiv e-prints},
         year = 2026,
        month = jul,
          eid = {arXiv:2607.24929},
        pages = {arXiv:2607.24929},
          doi = {10.48550/arXiv.2607.24929},
archivePrefix = {arXiv},
       eprint = {2607.24929},
 primaryClass = {astro-ph.GA},
       adsurl = {https://ui.adsabs.harvard.edu/abs/2026arXiv260724929L}
}

@ARTICLE{Chen2021,
       author = {{Chen}, Yangyao and {Mo}, H.~J. and {Li}, Cheng and {Wang}, Kai},
        title = "{How to empirically model star formation in dark matter haloes - I. Inferences about central galaxies from numerical simulations}",
      journal = {\mnras},
         year = 2021,
        month = jul,
       volume = {504},
       number = {4},
        pages = {4865-4884},
          doi = {10.1093/mnras/stab695},
archivePrefix = {arXiv},
       eprint = {2009.12467},
 primaryClass = {astro-ph.GA},
       adsurl = {https://ui.adsabs.harvard.edu/abs/2021MNRAS.504.4865C}
}

@ARTICLE{Mo2024,
       author = {{Mo}, Houjun and {Chen}, Yangyao and {Wang}, Huiyuan},
        title = "{A two-phase model of galaxy formation: I. The growth of galaxies and supermassive black holes}",
      journal = {\mnras},
         year = 2024,
        month = aug,
       volume = {532},
       number = {4},
        pages = {3808-3838},
          doi = {10.1093/mnras/stae1727},
archivePrefix = {arXiv},
       eprint = {2311.05030},
 primaryClass = {astro-ph.GA},
       adsurl = {https://ui.adsabs.harvard.edu/abs/2024MNRAS.532.3808M}
}

@ARTICLE{Li2015,
       author = {{Li}, Cheng and {Wang}, Enci and {Lin}, Lin and {Bershady}, Matthew A. and {Bundy}, Kevin and {Tremonti}, Christy A. and {Xiao}, Ting and {Yan}, Renbin and {Bizyaev}, Dmitry and {Blanton}, Michael and {Cales}, Sabrina and {Cherinka}, Brian and {Cheung}, Edmond and {Drory}, Niv and {Emsellem}, Eric and {Fu}, Hai and {Gelfand}, Joseph and {Law}, David R. and {Lin}, Lihwai and {MacDonald}, Nick and {Maraston}, Claudia and {Masters}, Karen L. and {Merrifield}, Michael R. and {Pan}, Kaike and {S{\'a}nchez}, S.~F. and {Schneider}, Donald P. and {Thomas}, Daniel and {Wake}, David and {Wang}, Lixin and {Weijmans}, Anne-Marie and {Wilkinson}, David and {Yoachim}, Peter and {Zhang}, Kai and {Zheng}, Tiantian},
        title = "{P-MaNGA: Gradients in Recent Star Formation Histories as Diagnostics for Galaxy Growth and Death}",
      journal = {\apj},
         year = 2015,
        month = may,
       volume = {804},
       number = {2},
          eid = {125},
        pages = {125},
          doi = {10.1088/0004-637X/804/2/125},
archivePrefix = {arXiv},
       eprint = {1502.07040},
 primaryClass = {astro-ph.GA},
       adsurl = {https://ui.adsabs.harvard.edu/abs/2015ApJ...804..125L}
}

@misc{jades_2023,
      title={Overview of the JWST Advanced Deep Extragalactic Survey (JADES)}, 
      author={Daniel J. Eisenstein and Chris Willott and Stacey Alberts and Santiago Arribas and Nina Bonaventura and Andrew J. Bunker and Alex J. Cameron and Stefano Carniani and Stephane Charlot and Emma Curtis-Lake and Francesco D'Eugenio and Ryan Endsley and Pierre Ferruit and Giovanna Giardino and Kevin Hainline and Ryan Hausen and Peter Jakobsen and Benjamin D. Johnson and Roberto Maiolino and Marcia Rieke and George Rieke and Hans-Walter Rix and Brant Robertson and Daniel P. Stark and Sandro Tacchella and Christina C. Williams and Christopher N. A. Willmer and William M. Baker and Stefi Baum and Rachana Bhatawdekar and Kristan Boyett and Zuyi Chen and Jacopo Chevallard and Chiara Circosta and Mirko Curti and A. Lola Danhaive and Christa DeCoursey and Anna de Graaff and Alan Dressler and Eiichi Egami and Jakob M. Helton and Raphael E. Hviding and Zhiyuan Ji and Gareth C. Jones and Nimisha Kumari and Nora Lützgendorf and Isaac Laseter and Tobias J. Looser and Jianwei Lyu and Michael V. Maseda and Erica Nelson and Eleonora Parlanti and Michele Perna and Dávid Puskás and Tim Rawle and Bruno Rodríguez Del Pino and Lester Sandles and Aayush Saxena and Jan Scholtz and Katherine Sharpe and Irene Shivaei and Maddie S. Silcock and Charlotte Simmonds and Maya Skarbinski and Renske Smit and Meredith Stone and Katherine A. Suess and Fengwu Sun and Mengtao Tang and Michael W. Topping and Hannah Übler and Natalia C. Villanueva and Imaan E. B. Wallace and Lily Whitler and Joris Witstok and Charity Woodrum},
      year={2023},
      eprint={2306.02465},
      archivePrefix={arXiv},
      primaryClass={astro-ph.GA},
      url={https://arxiv.org/abs/2306.02465}, 
}

@MISC{primer_2021,
       author = {{Dunlop}, James S. and {Abraham}, Roberto G. and {Ashby}, Matthew L.~N. and {Bagley}, Micaela and {Best}, Philip N. and {Bongiorno}, Angela and {Bouwens}, Rychard and {Bowler}, Rebecca A.~A. and {Brammer}, Gabriel and {Bremer}, Malcolm and {Calabro'}, Antonello and {Carnall}, Adam and {Castellano}, Marco and {Cirasuolo}, Michele and {Conselice}, Christopher and {Cullen}, Fergus and {Dave}, Romeel and {Dayal}, Pratika and {Dekel}, Avishai and {Dickinson}, Mark and {Duncan}, Kenneth James and {Elbaz}, David and {Ellis}, Richard S. and {Ferguson}, Harry C. and {Ferrara}, Andrea and {Finkelstein}, Steven L. and {Fontana}, Adriano and {Furlanetto}, Steven and {Fynbo}, Johan P.~U. and {Gallerani}, Simona and {Gardner}, Jonathan P. and {Giavalisco}, Mauro and {Grazian}, Andrea and {Grogin}, Norman and {Harikane}, Yuichi and {Hopkins}, Philip F. and {Ilbert}, Olivier and {Illingworth}, Garth D. and {Juneau}, Stephanie and {Jung}, Intae and {Kartaltepe}, Jeyhan and {Kassin}, Susan and {Kauffmann}, Olivier Benjamin and {Khochfar}, Sadegh and {Kirkpatrick}, Allison and {Kocevski}, Dale D. and {Koekemoer}, Anton M. and {Labbe}, Ivo and {Laporte}, Nicolas and {Larson}, Rebecca L. and {Lucas}, Ray A. and {Magee}, Daniel K. and {Mason}, Charlotte and {McCracken}, Henry Joy and {McLeod}, Derek and {McLure}, Ross and {Merlin}, Emiliano and {Mesinger}, Andrei and {Milvang-Jensen}, Bo and {Newman}, Jeffrey Allen and {Oesch}, Pascal and {Ouchi}, Masami and {Pacifici}, Camilla and {Papovich}, Casey and {Peacock}, John and {Peeples}, Molly and {Pentericci}, Laura and {Perez-Gonzalez}, Pablo G. and {Pirzkal}, Norbert and {Pope}, Alexandra and {Pye}, John P. and {Reddy}, Naveen A. and {Robertson}, Brant and {Salvato}, Mara and {Santini}, Paola and {Schaerer}, Daniel and {Shapley}, Alice E. and {Simons}, Raymond and {Smit}, Renske and {Smith}, Britton D. and {Snyder}, Greg and {Somerville}, Rachel S. and {Stanway}, Elizabeth R. and {Stefanon}, Mauro and {Tasca}, Lidia and {Tikkanen}, Tuomo and {Tresse}, Laurence and {Trump}, Jonathan R. and {Whitaker}, Katherine E. and {Wilkins}, Stephen Matthew and {Wright}, Gillian and {Wyithe}, J. Stuart B. and {van Dokkum}, Pieter and {van der Werf}, Paul},
        title = "{PRIMER: Public Release IMaging for Extragalactic Research}",
 howpublished = {JWST Proposal. Cycle 1, ID. \#1837},
         year = 2021,
        month = mar,
        pages = {1837},
       adsurl = {https://ui.adsabs.harvard.edu/abs/2021jwst.prop.1837D}
}

@article{Valentino_2023,
    doi = {10.3847/1538-4357/acbefa},
    url = {https://doi.org/10.3847/1538-4357/acbefa},
    year = {2023},
    month = {apr},
    publisher = {The American Astronomical Society},
    volume = {947},
    number = {1},
    pages = {20},
    author = {Valentino, Francesco and Brammer, Gabriel and Gould, Katriona M. L. and Kokorev, Vasily and Fujimoto, Seiji and Jespersen, Christian Kragh and Vijayan, Aswin P. and Weaver, John R. and Ito, Kei and Tanaka, Masayuki and Ilbert, Olivier and Magdis, Georgios E. and Whitaker, Katherine E. and Faisst, Andreas L. and Gallazzi, Anna and Gillman, Steven and Giménez-Arteaga, Clara and Gómez-Guijarro, Carlos and Kubo, Mariko and Heintz, Kasper E. and Hirschmann, Michaela and Oesch, Pascal and Onodera, Masato and Rizzo, Francesca and Lee, Minju and Strait, Victoria and Toft, Sune},
    journal = {The Astrophysical Journal}
    }

@ARTICLE{Peng_2002,
       author = {{Peng}, Chien Y. and {Ho}, Luis C. and {Impey}, Chris D. and {Rix}, Hans-Walter},
        title = "{Detailed Structural Decomposition of Galaxy Images}",
      journal = {\aj},
         year = 2002,
        month = jul,
       volume = {124},
       number = {1},
        pages = {266-293},
          doi = {10.1086/340952},
archivePrefix = {arXiv},
       eprint = {astro-ph/0204182},
 primaryClass = {astro-ph},
       adsurl = {https://ui.adsabs.harvard.edu/abs/2002AJ....124..266P}
}

@ARTICLE{Peng_2010,
       author = {{Peng}, Chien Y. and {Ho}, Luis C. and {Impey}, Chris D. and {Rix}, Hans-Walter},
        title = "{Detailed Decomposition of Galaxy Images. II. Beyond Axisymmetric Models}",
      journal = {\aj},
         year = 2010,
        month = jun,
       volume = {139},
       number = {6},
        pages = {2097-2129},
          doi = {10.1088/0004-6256/139/6/2097},
archivePrefix = {arXiv},
       eprint = {0912.0731},
 primaryClass = {astro-ph.CO},
       adsurl = {https://ui.adsabs.harvard.edu/abs/2010AJ....139.2097P}
}

@ARTICLE{Yao_2023,
       author = {{Yao}, Yao and {Song}, Jie and {Kong}, Xu and {Fang}, Guanwen and {Zhang}, Hong-Xin and {Chen}, Xinkai},
        title = "{Evolution of Nonparametric Morphology of Galaxies in the JWST CEERS Field at z ≃ 0.8-3.0}",
      journal = {\apj},
         year = 2023,
        month = sep,
       volume = {954},
       number = {2},
          eid = {113},
        pages = {113},
          doi = {10.3847/1538-4357/ace7b5},
archivePrefix = {arXiv},
       eprint = {2307.13975},
 primaryClass = {astro-ph.GA},
       adsurl = {https://ui.adsabs.harvard.edu/abs/2023ApJ...954..113Y}
}

@ARTICLE{WangWT_2019,
       author = {{Wang}, Wenting and {Han}, Jiaxin and {Sonnenfeld}, Alessandro and {Yasuda}, Naoki and {Li}, Xiangchong and {Jing}, Yipeng and {More}, Surhud and {Price}, Paul A. and {Lupton}, Robert and {Rykoff}, Eli S. and {Stark}, David V. and {Lan}, Ting-Wen and {Takada}, Masahiro and {Huang}, Song and {Luo}, Wentao and {Bahcall}, Neta A. and {Komiyama}, Yutaka},
        title = "{The stellar halo of isolated central galaxies in the Hyper Suprime-Cam imaging survey}",
      journal = {\mnras},
         year = 2019,
        month = aug,
       volume = {487},
       number = {2},
        pages = {1580-1606},
          doi = {10.1093/mnras/stz1339},
archivePrefix = {arXiv},
       eprint = {1811.04714},
 primaryClass = {astro-ph.GA},
       adsurl = {https://ui.adsabs.harvard.edu/abs/2019MNRAS.487.1580W}
}

@ARTICLE{Planck_2016,
       author = {{Planck Collaboration} and {Ade}, P.~A.~R. and {Aghanim}, N. and {Arnaud}, M. and {Ashdown}, M. and {Aumont}, J. and {Baccigalupi}, C. and {Banday}, A.~J. and {Barreiro}, R.~B. and {Bartlett}, J.~G. and {Bartolo}, N. and {Battaner}, E. and {Battye}, R. and {Benabed}, K. and {Beno{\^\i}t}, A. and {Benoit-L{\'e}vy}, A. and {Bernard}, J. -P. and {Bersanelli}, M. and {Bielewicz}, P. and {Bock}, J.~J. and {Bonaldi}, A. and {Bonavera}, L. and {Bond}, J.~R. and {Borrill}, J. and {Bouchet}, F.~R. and {Boulanger}, F. and {Bucher}, M. and {Burigana}, C. and {Butler}, R.~C. and {Calabrese}, E. and {Cardoso}, J. -F. and {Catalano}, A. and {Challinor}, A. and {Chamballu}, A. and {Chary}, R. -R. and {Chiang}, H.~C. and {Chluba}, J. and {Christensen}, P.~R. and {Church}, S. and {Clements}, D.~L. and {Colombi}, S. and {Colombo}, L.~P.~L. and {Combet}, C. and {Coulais}, A. and {Crill}, B.~P. and {Curto}, A. and {Cuttaia}, F. and {Danese}, L. and {Davies}, R.~D. and {Davis}, R.~J. and {de Bernardis}, P. and {de Rosa}, A. and {de Zotti}, G. and {Delabrouille}, J. and {D{\'e}sert}, F. -X. and {Di Valentino}, E. and {Dickinson}, C. and {Diego}, J.~M. and {Dolag}, K. and {Dole}, H. and {Donzelli}, S. and {Dor{\'e}}, O. and {Douspis}, M. and {Ducout}, A. and {Dunkley}, J. and {Dupac}, X. and {Efstathiou}, G. and {Elsner}, F. and {En{\ss}lin}, T.~A. and {Eriksen}, H.~K. and {Farhang}, M. and {Fergusson}, J. and {Finelli}, F. and {Forni}, O. and {Frailis}, M. and {Fraisse}, A.~A. and {Franceschi}, E. and {Frejsel}, A. and {Galeotta}, S. and {Galli}, S. and {Ganga}, K. and {Gauthier}, C. and {Gerbino}, M. and {Ghosh}, T. and {Giard}, M. and {Giraud-H{\'e}raud}, Y. and {Giusarma}, E. and {Gjerl{\o}w}, E. and {Gonz{\'a}lez-Nuevo}, J. and {G{\'o}rski}, K.~M. and {Gratton}, S. and {Gregorio}, A. and {Gruppuso}, A. and {Gudmundsson}, J.~E. and {Hamann}, J. and {Hansen}, F.~K. and {Hanson}, D. and {Harrison}, D.~L. and {Helou}, G. and {Henrot-Versill{\'e}}, S. and {Hern{\'a}ndez-Monteagudo}, C. and {Herranz}, D. and {Hildebrandt}, S.~R. and {Hivon}, E. and {Hobson}, M. and {Holmes}, W.~A. and {Hornstrup}, A. and {Hovest}, W. and {Huang}, Z. and {Huffenberger}, K.~M. and {Hurier}, G. and {Jaffe}, A.~H. and {Jaffe}, T.~R. and {Jones}, W.~C. and {Juvela}, M. and {Keih{\"a}nen}, E. and {Keskitalo}, R. and {Kisner}, T.~S. and {Kneissl}, R. and {Knoche}, J. and {Knox}, L. and {Kunz}, M. and {Kurki-Suonio}, H. and {Lagache}, G. and {L{\"a}hteenm{\"a}ki}, A. and {Lamarre}, J. -M. and {Lasenby}, A. and {Lattanzi}, M. and {Lawrence}, C.~R. and {Leahy}, J.~P. and {Leonardi}, R. and {Lesgourgues}, J. and {Levrier}, F. and {Lewis}, A. and {Liguori}, M. and {Lilje}, P.~B. and {Linden-V{\o}rnle}, M. and {L{\'o}pez-Caniego}, M. and {Lubin}, P.~M. and {Mac{\'\i}as-P{\'e}rez}, J.~F. and {Maggio}, G. and {Maino}, D. and {Mandolesi}, N. and {Mangilli}, A. and {Marchini}, A. and {Maris}, M. and {Martin}, P.~G. and {Martinelli}, M. and {Mart{\'\i}nez-Gonz{\'a}lez}, E. and {Masi}, S. and {Matarrese}, S. and {McGehee}, P. and {Meinhold}, P.~R. and {Melchiorri}, A. and {Melin}, J. -B. and {Mendes}, L. and {Mennella}, A. and {Migliaccio}, M. and {Millea}, M. and {Mitra}, S. and {Miville-Desch{\^e}nes}, M. -A. and {Moneti}, A. and {Montier}, L. and {Morgante}, G. and {Mortlock}, D. and {Moss}, A. and {Munshi}, D. and {Murphy}, J.~A. and {Naselsky}, P. and {Nati}, F. and {Natoli}, P. and {Netterfield}, C.~B. and {N{\o}rgaard-Nielsen}, H.~U. and {Noviello}, F. and {Novikov}, D. and {Novikov}, I. and {Oxborrow}, C.~A. and {Paci}, F. and {Pagano}, L. and {Pajot}, F. and {Paladini}, R. and {Paoletti}, D. and {Partridge}, B. and {Pasian}, F. and {Patanchon}, G. and {Pearson}, T.~J. and {Perdereau}, O. and {Perotto}, L. and {Perrotta}, F. and {Pettorino}, V. and {Piacentini}, F. and {Piat}, M. and {Pierpaoli}, E. and {Pietrobon}, D. and {Plaszczynski}, S. and {Pointecouteau}, E. and {Polenta}, G. and {Popa}, L. and {Pratt}, G.~W. and {Pr{\'e}zeau}, G.},
        title = "{Planck 2015 results. XIII. Cosmological parameters}",
      journal = {\aap},
         year = 2016,
        month = sep,
       volume = {594},
          eid = {A13},
        pages = {A13},
          doi = {10.1051/0004-6361/201525830},
archivePrefix = {arXiv},
       eprint = {1502.01589},
 primaryClass = {astro-ph.CO},
       adsurl = {https://ui.adsabs.harvard.edu/abs/2016A&A...594A..13P}
}

@ARTICLE{Williams_2025,
       author = {{Williams}, Devin J. and {Damjanov}, Ivana and {Sawicki}, Marcin and {Souchereau}, Harrison and {Chen}, Lingjian and {Desprez}, Guillaume and {George}, Angelo and {Annunziatella}, Marianna and {Arnouts}, St{\'e}phane and {Gwyn}, Stephen and {Marchesini}, Danilo and {Sajina}, Anna},
        title = "{The Growth of Galaxy Stellar Haloes over 0.2 {\ensuremath{\leq}} z {\ensuremath{\leq}} 1.1}",
      journal = {\apj},
         year = 2025,
        month = aug,
       volume = {989},
       number = {1},
          eid = {107},
        pages = {107},
          doi = {10.3847/1538-4357/ade9a8},
archivePrefix = {arXiv},
       eprint = {2412.03662},
 primaryClass = {astro-ph.GA},
       adsurl = {https://ui.adsabs.harvard.edu/abs/2025ApJ...989..107W}
}

@ARTICLE{Tao_2026,
       author = {{Tao}, Bojun and {Zhang}, Hong-Xin and {Wang}, Wenting and {Wang}, Enci and {Chen}, Guangwen and {Wang}, Huiyuan and {Chen}, Lijun and {Chen}, Qian-Hui and {Huang}, Song and {Kong}, Xu and {Rong}, Yu},
        title = "{The First Systematic Survey of Stellar Halos in High-inclination Galaxies Reveals Unusually Quiescent Merger Histories of Nearby Galaxies}",
      journal = {\apjs},
         year = 2026,
        month = may,
       volume = {284},
       number = {1},
          eid = {31},
        pages = {31},
          doi = {10.3847/1538-4365/ae48ec},
archivePrefix = {arXiv},
       eprint = {2602.20258},
 primaryClass = {astro-ph.GA},
       adsurl = {https://ui.adsabs.harvard.edu/abs/2026ApJS..284...31T}
}

@ARTICLE{Keller_2022,
       author = {{Keller}, B.~W.},
        title = "{Where Did the Outskirts Go? Outer Stellar Halos as a Sensitive Probe of Supernova Feedback}",
      journal = {\apj},
         year = 2022,
        month = nov,
       volume = {939},
       number = {1},
          eid = {4},
        pages = {4},
          doi = {10.3847/1538-4357/ac92fe},
archivePrefix = {arXiv},
       eprint = {2103.09833},
 primaryClass = {astro-ph.GA},
       adsurl = {https://ui.adsabs.harvard.edu/abs/2022ApJ...939....4K}
}

@ARTICLE{Elias_2018,
       author = {{Elias}, Lydia M. and {Sales}, Laura V. and {Creasey}, Peter and {Cooper}, Michael C. and {Bullock}, James S. and {Rich}, R. Michael and {Hernquist}, Lars},
        title = "{Stellar halos in Illustris: probing the histories of Milky Way-mass galaxies}",
      journal = {\mnras},
         year = 2018,
        month = sep,
       volume = {479},
       number = {3},
        pages = {4004-4016},
          doi = {10.1093/mnras/sty1718},
archivePrefix = {arXiv},
       eprint = {1801.07273},
 primaryClass = {astro-ph.GA},
       adsurl = {https://ui.adsabs.harvard.edu/abs/2018MNRAS.479.4004E}
}

@ARTICLE{Merritt_2020,
       author = {{Merritt}, Allison and {Pillepich}, Annalisa and {van Dokkum}, Pieter and {Nelson}, Dylan and {Hernquist}, Lars and {Marinacci}, Federico and {Vogelsberger}, Mark},
        title = "{A missing outskirts problem? Comparisons between stellar haloes in the Dragonfly Nearby Galaxies Survey and the TNG100 simulation}",
      journal = {\mnras},
         year = 2020,
        month = jul,
       volume = {495},
       number = {4},
        pages = {4570-4604},
          doi = {10.1093/mnras/staa1164},
archivePrefix = {arXiv},
       eprint = {2004.11402},
 primaryClass = {astro-ph.GA},
       adsurl = {https://ui.adsabs.harvard.edu/abs/2020MNRAS.495.4570M}
}

@ARTICLE{Pillepich_2015,
       author = {{Pillepich}, Annalisa and {Madau}, Piero and {Mayer}, Lucio},
        title = "{Building Late-type Spiral Galaxies by In-situ and Ex-situ Star Formation}",
      journal = {\apj},
         year = 2015,
        month = feb,
       volume = {799},
       number = {2},
          eid = {184},
        pages = {184},
          doi = {10.1088/0004-637X/799/2/184},
archivePrefix = {arXiv},
       eprint = {1407.7855},
 primaryClass = {astro-ph.GA},
       adsurl = {https://ui.adsabs.harvard.edu/abs/2015ApJ...799..184P}
}

@ARTICLE{Cannarozzo_2023,
       author = {{Cannarozzo}, Carlo and {Leauthaud}, Alexie and {Oyarz{\'u}n}, Grecco A. and {Nipoti}, Carlo and {Diemer}, Benedikt and {Huang}, Song and {Rodriguez-Gomez}, Vicente and {Sonnenfeld}, Alessandro and {Bundy}, Kevin},
        title = "{The contribution of in situ and ex situ star formation in early-type galaxies: MaNGA versus IllustrisTNG}",
      journal = {\mnras},
         year = 2023,
        month = apr,
       volume = {520},
       number = {4},
        pages = {5651-5670},
          doi = {10.1093/mnras/stac3023},
archivePrefix = {arXiv},
       eprint = {2210.08109},
 primaryClass = {astro-ph.GA},
       adsurl = {https://ui.adsabs.harvard.edu/abs/2023MNRAS.520.5651C}
}

@ARTICLE{RG_2016,
       author = {{Rodriguez-Gomez}, Vicente and {Pillepich}, Annalisa and {Sales}, Laura V. and {Genel}, Shy and {Vogelsberger}, Mark and {Zhu}, Qirong and {Wellons}, Sarah and {Nelson}, Dylan and {Torrey}, Paul and {Springel}, Volker and {Ma}, Chung-Pei and {Hernquist}, Lars},
        title = "{The stellar mass assembly of galaxies in the Illustris simulation: growth by mergers and the spatial distribution of accreted stars}",
      journal = {\mnras},
         year = 2016,
        month = may,
       volume = {458},
       number = {3},
        pages = {2371-2390},
          doi = {10.1093/mnras/stw456},
archivePrefix = {arXiv},
       eprint = {1511.08804},
 primaryClass = {astro-ph.GA},
       adsurl = {https://ui.adsabs.harvard.edu/abs/2016MNRAS.458.2371R}
}

@ARTICLE{Ownsworth_2014,
       author = {{Ownsworth}, Jamie R. and {Conselice}, Christopher J. and {Mortlock}, Alice and {Hartley}, William G. and {Almaini}, Omar and {Duncan}, Ken and {Mundy}, Carl J.},
        title = "{Minor versus major mergers: the stellar mass growth of massive galaxies from z = 3 using number density selection techniques}",
      journal = {\mnras},
         year = 2014,
        month = dec,
       volume = {445},
       number = {3},
        pages = {2198-2213},
          doi = {10.1093/mnras/stu1802},
archivePrefix = {arXiv},
       eprint = {1409.1582},
 primaryClass = {astro-ph.GA},
       adsurl = {https://ui.adsabs.harvard.edu/abs/2014MNRAS.445.2198O}
}

@ARTICLE{Conselice_2022,
       author = {{Conselice}, Christopher J. and {Mundy}, Carl J. and {Ferreira}, Leonardo and {Duncan}, Kenneth},
        title = "{A Direct Measurement of Galaxy Major and Minor Merger Rates and Stellar Mass Accretion Histories at Z < 3 Using Galaxy Pairs in the REFINE Survey}",
      journal = {\apj},
         year = 2022,
        month = dec,
       volume = {940},
       number = {2},
          eid = {168},
        pages = {168},
          doi = {10.3847/1538-4357/ac9b1a},
archivePrefix = {arXiv},
       eprint = {2207.03984},
 primaryClass = {astro-ph.GA},
       adsurl = {https://ui.adsabs.harvard.edu/abs/2022ApJ...940..168C}
}

@ARTICLE{Duan_2026,
       author = {{Duan}, Qiao and {Conselice}, Christopher J. and {Harvey}, Thomas and {Li}, Qiong and {Austin}, Duncan and {Adams}, Nathan J. and {Ferreira}, Leonardo and {Duncan}, Kenneth J. and {Trussler}, James and {Pascalau}, Robert G. and {Windhorst}, Rogier A. and {Holwerda}, Benne W. and {Broadhurst}, Thomas J. and {Coe}, Dan and {Cohen}, Seth H. and {Du}, Xiaojing and {Driver}, Simon P. and {Frye}, Brenda and {Grogin}, Norman A. and {Hathi}, Nimish P. and {Jansen}, Rolf A. and {Koekemoer}, Anton M. and {Marshall}, Madeline A. and {Nonino}, Mario and {Ortiz}, III, Rafael and {Pirzkal}, Nor and {Robotham}, Aaron and {Ryan}, Jr., Russell E. and {Summers}, Jake and {D'Silva}, Jordan C.~J. and {Willmer}, Christopher N.~A. and {Yan}, Haojing},
        title = "{Galaxy mergers in the epoch of reionization ─ II. Merger-triggered star formation and AGN activities at z = 4.5{\ensuremath{-}}8.5}",
      journal = {\mnras},
         year = 2026,
        month = mar,
       volume = {546},
       number = {3},
          eid = {stag008},
        pages = {stag008},
          doi = {10.1093/mnras/stag008},
archivePrefix = {arXiv},
       eprint = {2411.04944},
 primaryClass = {astro-ph.GA},
       adsurl = {https://ui.adsabs.harvard.edu/abs/2026MNRAS.546ag008D}
}

@ARTICLE{Chen_2024,
       author = {{Chen}, Yangyao and {Mo}, Houjun and {Wang}, Huiyuan},
        title = "{A two-phase model of galaxy formation - II. The size-mass relation of dynamically hot galaxies}",
      journal = {\mnras},
         year = 2024,
        month = aug,
       volume = {532},
       number = {4},
        pages = {4340-4349},
          doi = {10.1093/mnras/stae1757},
archivePrefix = {arXiv},
       eprint = {2311.11713},
 primaryClass = {astro-ph.GA},
       adsurl = {https://ui.adsabs.harvard.edu/abs/2024MNRAS.532.4340C}
}

@ARTICLE{Tripodi_2024,
       author = {{Tripodi}, Roberta and {D'Eugenio}, Francesco and {Maiolino}, Roberto and {Curti}, Mirko and {Scholtz}, Jan and {Tacchella}, Sandro and {Marconcini}, Cosimo and {Bunker}, Andrew J. and {Trussler}, James A.~A. and {Cameron}, Alex J. and {Arribas}, Santiago and {Baker}, William M. and {Brada{\v{c}}}, Maru{\v{s}}a and {Carniani}, Stefano and {Charlot}, St{\'e}phane and {Ji}, Xihan and {Ji}, Zhiyuan and {Robertson}, Brant and {{\"U}bler}, Hannah and {Venturi}, Giacomo and {Willmer}, Christopher N.~A. and {Witstok}, Joris},
        title = "{Spatially resolved emission lines in galaxies at 4 {\ensuremath{\leq}} z < 10 from the JADES survey: Evidence for enhanced central star formation}",
      journal = {\aap},
         year = 2024,
        month = dec,
       volume = {692},
          eid = {A184},
        pages = {A184},
          doi = {10.1051/0004-6361/202449980},
archivePrefix = {arXiv},
       eprint = {2403.08431},
 primaryClass = {astro-ph.GA},
       adsurl = {https://ui.adsabs.harvard.edu/abs/2024A&A...692A.184T}
}

@ARTICLE{Jia_2024,
       author = {{Jia}, Cheng and {Wang}, Enci and {Wang}, Huiyuan and {Li}, Hui and {Yao}, Yao and {Song}, Jie and {Zhang}, Hongxin and {Rong}, Yu and {Chen}, Yangyao and {Yu}, Haoran and {Chen}, Zeyu and {Li}, Haixin and {Ma}, Chengyu and {Kong}, Xu},
        title = "{Size Growth on Short Timescales of Star-forming Galaxies: Insights from Size Variation with Rest-frame Wavelength with JADES}",
      journal = {\apj},
         year = 2024,
        month = dec,
       volume = {977},
       number = {2},
          eid = {165},
        pages = {165},
          doi = {10.3847/1538-4357/ad919a},
archivePrefix = {arXiv},
       eprint = {2411.07458},
 primaryClass = {astro-ph.GA},
       adsurl = {https://ui.adsabs.harvard.edu/abs/2024ApJ...977..165J}
}

@ARTICLE{Yang_2025,
       author = {{Yang}, Lilan and {Kartaltepe}, Jeyhan S. and {Franco}, Maximilien and {Ding}, Xuheng and {Achenbach}, Mark J. and {Arango-Toro}, Rafael C. and {Casey}, Caitlin M. and {Drakos}, Nicole E. and {Faisst}, Andreas L. and {Gillman}, Steven and {Gozaliasl}, Ghassem and {Huertas-Company}, Marc and {Jin}, Shuowen and {Liu}, Daizhong and {Magdis}, Georgios and {Massey}, Richard and {Silverman}, John D. and {Tanaka}, Takumi S. and {Yu}, Si-Yue and {Akins}, Hollis B. and {Allen}, Natalie and {Ilbert}, Olivier and {Koekemoer}, Anton M. and {McCracken}, Henry Joy and {Paquereau}, Louise and {Rhodes}, Jason and {Robertson}, Brant E. and {Shuntov}, Marko and {Toft}, Sune},
        title = "{COSMOS-Web: Unraveling the Evolution of Galaxy Size and Related Properties at 2 < z < 10}",
      journal = {\apjs},
         year = 2025,
        month = dec,
       volume = {281},
       number = {2},
          eid = {68},
        pages = {68},
          doi = {10.3847/1538-4365/ae0e1b},
archivePrefix = {arXiv},
       eprint = {2504.07185},
 primaryClass = {astro-ph.GA},
       adsurl = {https://ui.adsabs.harvard.edu/abs/2025ApJS..281...68Y}
}

@article{Ormerod_2023,
   title={EPOCHS VI: the size and shape evolution of galaxies since <i>z</i> ∼ 8 with <i>JWST</i> Observations},
   volume={527},
   ISSN={1365-2966},
   url={http://dx.doi.org/10.1093/mnras/stad3597},
   DOI={10.1093/mnras/stad3597},
   number={3},
   journal={Monthly Notices of the Royal Astronomical Society},
   publisher={Oxford University Press (OUP)},
   author={Ormerod, K and Conselice, C J and Adams, N J and Harvey, T and Austin, D and Trussler, J and Ferreira, L and Caruana, J and Lucatelli, G and Li, Q and Roper, W J},
   year={2023},
   month=Nov, pages={6110–6125} }

@ARTICLE{Ward_2024,
       author = {{Ward}, Ethan and {de la Vega}, Alexander and {Mobasher}, Bahram and {McGrath}, Elizabeth J. and {Iyer}, Kartheik G. and {Calabr{\`o}}, Antonello and {Costantin}, Luca and {Dickinson}, Mark and {Holwerda}, Benne W. and {Huertas-Company}, Marc and {Hirschmann}, Michaela and {Lucas}, Ray A. and {Pandya}, Viraj and {Wilkins}, Stephen M. and {Yung}, L.~Y. Aaron and {Arrabal Haro}, Pablo and {Bagley}, Micaela B. and {Finkelstein}, Steven L. and {Kartaltepe}, Jeyhan S. and {Koekemoer}, Anton M. and {Papovich}, Casey and {Pirzkal}, Nor},
        title = "{Evolution of the Size─Mass Relation of Star-forming Galaxies Since z = 5.5 Revealed by CEERS}",
      journal = {\apj},
         year = 2024,
        month = feb,
       volume = {962},
       number = {2},
          eid = {176},
        pages = {176},
          doi = {10.3847/1538-4357/ad20ed},
archivePrefix = {arXiv},
       eprint = {2311.02162},
 primaryClass = {astro-ph.GA},
       adsurl = {https://ui.adsabs.harvard.edu/abs/2024ApJ...962..176W}
}

@ARTICLE{Morishita_2024,
       author = {{Morishita}, Takahiro and {Stiavelli}, Massimo and {Chary}, Ranga-Ram and {Trenti}, Michele and {Bergamini}, Pietro and {Chiaberge}, Marco and {Leethochawalit}, Nicha and {Roberts-Borsani}, Guido and {Shen}, Xuejian and {Treu}, Tommaso},
        title = "{Enhanced Subkiloparsec-scale Star Formation: Results from a JWST Size Analysis of 341 Galaxies at 5 < z < 14}",
      journal = {\apj},
         year = 2024,
        month = mar,
       volume = {963},
       number = {1},
          eid = {9},
        pages = {9},
          doi = {10.3847/1538-4357/ad1404},
archivePrefix = {arXiv},
       eprint = {2308.05018},
 primaryClass = {astro-ph.GA},
       adsurl = {https://ui.adsabs.harvard.edu/abs/2024ApJ...963....9M}
}

@ARTICLE{Allen_2025,
       author = {{Allen}, Natalie and {Oesch}, Pascal A. and {Toft}, Sune and {Matharu}, Jasleen and {McPartland}, Conor J.~R. and {Weibel}, Andrea and {Brammer}, Gabe and {Bowler}, Rebecca A.~A. and {Ito}, Kei and {Gottumukkala}, Rashmi and {Rizzo}, Francesca and {Valentino}, Francesco and {Varadaraj}, Rohan G. and {Weaver}, John R. and {Whitaker}, Katherine E.},
        title = "{Galaxy size and mass build-up in the first 2 Gyr of cosmic history from multi-wavelength JWST NIRCam imaging}",
      journal = {\aap},
         year = 2025,
        month = jun,
       volume = {698},
          eid = {A30},
        pages = {A30},
          doi = {10.1051/0004-6361/202452690},
archivePrefix = {arXiv},
       eprint = {2410.16354},
 primaryClass = {astro-ph.GA},
       adsurl = {https://ui.adsabs.harvard.edu/abs/2025A&A...698A..30A}
}

@ARTICLE{Zhu_2022,
       author = {{Zhu}, Ling and {Pillepich}, Annalisa and {van de Ven}, Glenn and {Leaman}, Ryan and {Hernquist}, Lars and {Nelson}, Dylan and {Pakmor}, Ruediger and {Vogelsberger}, Mark and {Zhang}, Le},
        title = "{Mass of the dynamically hot inner stellar halo predicts the ancient accreted stellar mass}",
      journal = {\aap},
         year = 2022,
        month = apr,
       volume = {660},
          eid = {A20},
        pages = {A20},
          doi = {10.1051/0004-6361/202142496},
archivePrefix = {arXiv},
       eprint = {2110.13172},
 primaryClass = {astro-ph.GA},
       adsurl = {https://ui.adsabs.harvard.edu/abs/2022A&A...660A..20Z}
}

@ARTICLE{Di-Matteo_2019,
       author = {{Di Matteo}, P. and {Haywood}, M. and {Lehnert}, M.~D. and {Katz}, D. and {Khoperskov}, S. and {Snaith}, O.~N. and {G{\'o}mez}, A. and {Robichon}, N.},
        title = "{The Milky Way has no in-situ halo other than the heated thick disc. Composition of the stellar halo and age-dating the last significant merger with Gaia DR2 and APOGEE}",
      journal = {\aap},
         year = 2019,
        month = dec,
       volume = {632},
          eid = {A4},
        pages = {A4},
          doi = {10.1051/0004-6361/201834929},
archivePrefix = {arXiv},
       eprint = {1812.08232},
 primaryClass = {astro-ph.GA},
       adsurl = {https://ui.adsabs.harvard.edu/abs/2019A&A...632A...4D}
}

@ARTICLE{Gozman_2023,
       author = {{Gozman}, Katya and {Bell}, Eric F. and {Smercina}, Adam and {Price}, Paul and {Bailin}, Jeremy and {de Jong}, Roelof S. and {D'Souza}, Richard and {Jang}, In Sung and {Monachesi}, Antonela and {Slater}, Colin},
        title = "{Saying Hallo to M94's Stellar Halo: Investigating the Accretion History of the Largest Pseudobulge Host in the Local Universe}",
      journal = {\apj},
         year = 2023,
        month = apr,
       volume = {947},
       number = {1},
          eid = {21},
        pages = {21},
          doi = {10.3847/1538-4357/acbe3a},
archivePrefix = {arXiv},
       eprint = {2304.08436},
 primaryClass = {astro-ph.GA},
       adsurl = {https://ui.adsabs.harvard.edu/abs/2023ApJ...947...21G}
}

@ARTICLE{Helmi_2020,
       author = {{Helmi}, Amina},
        title = "{Streams, Substructures, and the Early History of the Milky Way}",
      journal = {\araa},
         year = 2020,
        month = aug,
       volume = {58},
        pages = {205-256},
          doi = {10.1146/annurev-astro-032620-021917},
archivePrefix = {arXiv},
       eprint = {2002.04340},
 primaryClass = {astro-ph.GA},
       adsurl = {https://ui.adsabs.harvard.edu/abs/2020ARA&A..58..205H}
}

@ARTICLE{Gonzalez-Jara_2025,
       author = {{Gonzalez-Jara}, Jenny and {Tissera}, Patricia B. and {Monachesi}, Antonela and {Sillero}, Emanuel and {Pallero}, Diego and {Pedrosa}, Susana and {Tau}, Elisa A. and {Tapia-Contreras}, Brian and {Bignone}, Lucas},
        title = "{Unveiling the formation channels of stellar halos through their chemical fingerprints}",
      journal = {\aap},
         year = 2025,
        month = jan,
       volume = {693},
          eid = {A282},
        pages = {A282},
          doi = {10.1051/0004-6361/202452639},
archivePrefix = {arXiv},
       eprint = {2412.13483},
 primaryClass = {astro-ph.GA},
       adsurl = {https://ui.adsabs.harvard.edu/abs/2025A&A...693A.282G}
}

@ARTICLE{Bullock_2005,
       author = {{Bullock}, James S. and {Johnston}, Kathryn V.},
        title = "{Tracing Galaxy Formation with Stellar Halos. I. Methods}",
      journal = {\apj},
         year = 2005,
        month = dec,
       volume = {635},
       number = {2},
        pages = {931-949},
          doi = {10.1086/497422},
archivePrefix = {arXiv},
       eprint = {astro-ph/0506467},
 primaryClass = {astro-ph},
       adsurl = {https://ui.adsabs.harvard.edu/abs/2005ApJ...635..931B}
}

@ARTICLE{Wright_2024,
       author = {{Wright}, Anna C. and {Tumlinson}, Jason and {Peeples}, Molly S. and {O'Shea}, Brian W. and {Lochhaas}, Cassandra and {Corlies}, Lauren and {Smith}, Britton D. and {Binh}, Nguyen and {Augustin}, Ramona and {Simons}, Raymond C.},
        title = "{Figuring Out Gas and Galaxies in Enzo (FOGGIE). VII. The (Dis)assembly of Stellar Halos}",
      journal = {\apj},
         year = 2024,
        month = jul,
       volume = {970},
       number = {1},
          eid = {70},
        pages = {70},
          doi = {10.3847/1538-4357/ad49a3},
archivePrefix = {arXiv},
       eprint = {2309.10039},
 primaryClass = {astro-ph.GA},
       adsurl = {https://ui.adsabs.harvard.edu/abs/2024ApJ...970...70W}
}

@ARTICLE{Pulsoni_2021,
       author = {{Pulsoni}, C. and {Gerhard}, O. and {Arnaboldi}, M. and {Pillepich}, A. and {Rodriguez-Gomez}, V. and {Nelson}, D. and {Hernquist}, L. and {Springel}, V.},
        title = "{The stellar halos of ETGs in the IllustrisTNG simulations. II. Accretion, merger history, and dark halo connection}",
      journal = {\aap},
         year = 2021,
        month = mar,
       volume = {647},
          eid = {A95},
        pages = {A95},
          doi = {10.1051/0004-6361/202039166},
archivePrefix = {arXiv},
       eprint = {2009.01823},
 primaryClass = {astro-ph.GA},
       adsurl = {https://ui.adsabs.harvard.edu/abs/2021A&A...647A..95P}
}

@ARTICLE{Dodd_2025,
       author = {{Dodd}, Emma and {Matsuno}, Tadafumi and {Helmi}, Amina and {Balbinot}, Eduardo and {Callingham}, Thomas M. and {Starkenburg}, Else and {Woudenberg}, Hanneke C. and {Ruiz-Lara}, Tom{\'a}s},
        title = "{Chemical characterisation of small substructures in the local stellar halo}",
      journal = {\aap},
         year = 2025,
        month = aug,
       volume = {700},
          eid = {A154},
        pages = {A154},
          doi = {10.1051/0004-6361/202554252},
archivePrefix = {arXiv},
       eprint = {2502.17353},
 primaryClass = {astro-ph.GA},
       adsurl = {https://ui.adsabs.harvard.edu/abs/2025A&A...700A.154D}
}

@ARTICLE{Ogami_2025,
       author = {{Ogami}, Itsuki and {Tanaka}, Mikito and {Komiyama}, Yutaka and {Chiba}, Masashi and {Guhathakurta}, Puragra and {Kirby}, Evan N. and {Wyse}, Rosemary F.~G. and {Filion}, Carrie and {Gilbert}, Karoline M. and {Escala}, Ivanna and {Mori}, Masao and {Kirihara}, Takanobu and {Tanaka}, Masayuki and {Ishigaki}, Miho N. and {Hayashi}, Kohei and {Lee}, Myun Gyoon and {Sharma}, Sanjib and {Kalirai}, Jason S. and {Lupton}, Robert H.},
        title = "{The structure of the stellar halo of the Andromeda galaxy explored with the NB515 for Subaru/HSC - I. New insights on the stellar halo up to 120 kpc}",
      journal = {\mnras},
         year = 2025,
        month = jan,
       volume = {536},
       number = {1},
        pages = {530-553},
          doi = {10.1093/mnras/stae2527},
archivePrefix = {arXiv},
       eprint = {2401.00668},
 primaryClass = {astro-ph.GA},
       adsurl = {https://ui.adsabs.harvard.edu/abs/2025MNRAS.536..530O}
}

@misc{Li_2026,
      title={The Milky Way stellar halo is twisted and doubly broken: insights from DESI DR2 Milky Way Survey observation}, 
      author={Songting Li and Wenting Wang and Sergey E. Koposov and Joao A. S. Amarante and Alis J. Deason and Nathan R. Sandford and Ting S. Li and Gustavo E. Medina and Jaxin Han and Monica Valluri and Oleg Y. Gnedin and Namitha Kizhuprakkat and Andrew P. Cooper and Leandro Beraldo e Silva and Carlos Frenk and Raymond G. Carlberg and Mika Lambert and Tian Qiu and Jessica Nicole Aguilar and Steven Ahlen and Davide Bianchi and David Brooks and Todd Claybaugh and Axel de la Macorra and Peter Doel and Jaime E. Forero-Romero and Enrique Gaztanaga and Satya Gontcho A Gontcho and Gaston Gutierrez and Dick Joyce and Robert Kehoe and Anthony Kremin and Claire Lamman and Martin Landriau and Laurent Le Guillou and Ramon Miquel and Will Percival and Francisco Prada and Ignasi Perez-Rafols and Graziano Rossi and Eusebio Sanchez and David Schlegel and Ray Sharples and Joseph Harry Silber and David Sprayberry and Gregory Tarle and Benjamin Alan Weaver and Hu Zou},
      year={2026},
      eprint={2512.01350},
      archivePrefix={arXiv},
      primaryClass={astro-ph.GA},
      url={https://arxiv.org/abs/2512.01350}, 
}

@misc{Suzuki_2026,
      title={The Milky Way Tomography with Subaru Hyper Suprime-Cam. II. Global halo structure}, 
      author={Yoshihisa Suzuki and Masashi Chiba and Rosemary F. G. Wyse and Shunichi Horigome},
      year={2026},
      eprint={2604.10280},
      archivePrefix={arXiv},
      primaryClass={astro-ph.GA},
      url={https://arxiv.org/abs/2604.10280}, 
}

@ARTICLE{Xu_2025,
       author = {{Xu}, Shuo and {Huang}, Song and {Leauthaud}, Alexie and {Diemer}, Benedikt and {Leidig}, Katya and {Cannarozzo}, Carlo and {Zhou}, Conghao},
        title = "{The Outskirt Stellar Mass of Low-redshift Massive Galaxies Is an Excellent Halo Mass Proxy in Illustris/IllustrisTNG Simulations}",
      journal = {\apj},
         year = 2025,
        month = sep,
       volume = {990},
       number = {2},
          eid = {127},
        pages = {127},
          doi = {10.3847/1538-4357/adeca0},
archivePrefix = {arXiv},
       eprint = {2412.03406},
 primaryClass = {astro-ph.GA},
       adsurl = {https://ui.adsabs.harvard.edu/abs/2025ApJ...990..127X}
}

@article{Cooper_2025,
   title={Simulations of the accreted stellar haloes of low-mass field galaxies},
   volume={540},
   ISSN={1365-2966},
   url={http://dx.doi.org/10.1093/mnras/staf833},
   DOI={10.1093/mnras/staf833},
   number={3},
   journal={Monthly Notices of the Royal Astronomical Society},
   publisher={Oxford University Press (OUP)},
   author={Cooper, Andrew P and Frenk, Carlos S and Hellwing, Wojciech A and Bose, Sownak},
   year={2025},
   month=May, pages={2049–2080} }

@ARTICLE{Morrison_2000,
       author = {{Morrison}, Heather L. and {Mateo}, Mario and {Olszewski}, Edward W. and {Harding}, Paul and {Dohm-Palmer}, R.~C. and {Freeman}, Kenneth C. and {Norris}, John E. and {Morita}, Miwa},
        title = "{Mapping the Galactic Halo. I. The ``Spaghetti'' Survey}",
      journal = {\aj},
         year = 2000,
        month = may,
       volume = {119},
       number = {5},
        pages = {2254-2273},
          doi = {10.1086/301357},
archivePrefix = {arXiv},
       eprint = {astro-ph/0001492},
 primaryClass = {astro-ph},
       adsurl = {https://ui.adsabs.harvard.edu/abs/2000AJ....119.2254M}
}

@article{Tumlinson_2009,
   title={CHEMICAL EVOLUTION IN HIERARCHICAL MODELS OF COSMIC STRUCTURE. II. THE FORMATION OF THE MILKY WAY STELLAR HALO AND THE DISTRIBUTION OF THE OLDEST STARS},
   volume={708},
   ISSN={1538-4357},
   url={http://dx.doi.org/10.1088/0004-637X/708/2/1398},
   DOI={10.1088/0004-637x/708/2/1398},
   number={2},
   journal={The Astrophysical Journal},
   publisher={American Astronomical Society},
   author={Tumlinson, Jason},
   year={2009},
   month=Dec, pages={1398–1418} }

@ARTICLE{Bell_2008,
       author = {{Bell}, Eric F. and {Zucker}, Daniel B. and {Belokurov}, Vasily and {Sharma}, Sanjib and {Johnston}, Kathryn V. and {Bullock}, James S. and {Hogg}, David W. and {Jahnke}, Knud and {de Jong}, Jelte T.~A. and {Beers}, Timothy C. and {Evans}, N.~W. and {Grebel}, Eva K. and {Ivezi{\'c}}, {\v{Z}}eljko and {Koposov}, Sergey E. and {Rix}, Hans-Walter and {Schneider}, Donald P. and {Steinmetz}, Matthias and {Zolotov}, Adi},
        title = "{The Accretion Origin of the Milky Way's Stellar Halo}",
      journal = {\apj},
         year = 2008,
        month = jun,
       volume = {680},
       number = {1},
        pages = {295-311},
          doi = {10.1086/588032},
archivePrefix = {arXiv},
       eprint = {0706.0004},
 primaryClass = {astro-ph},
       adsurl = {https://ui.adsabs.harvard.edu/abs/2008ApJ...680..295B}
}

@ARTICLE{Moster_2013,
       author = {{Moster}, Benjamin P. and {Naab}, Thorsten and {White}, Simon D.~M.},
        title = "{Galactic star formation and accretion histories from matching galaxies to dark matter haloes}",
      journal = {\mnras},
         year = 2013,
        month = feb,
       volume = {428},
       number = {4},
        pages = {3121-3138},
          doi = {10.1093/mnras/sts261},
archivePrefix = {arXiv},
       eprint = {1205.5807},
 primaryClass = {astro-ph.CO},
       adsurl = {https://ui.adsabs.harvard.edu/abs/2013MNRAS.428.3121M}
}

@article{Kauffmann_2003,
    author = {Kauffmann, Guinevere and Heckman, Timothy M. and White, Simon D. M. and Charlot, Stéphane and Tremonti, Christy and Brinchmann, Jarle and Bruzual, Gustavo and Peng, Eric W. and Seibert, Mark and Bernardi, Mariangela and Blanton, Michael and Brinkmann, Jon and Castander, Francisco and Csábai, Istvan and Fukugita, Masataka and Ivezic, Zeljko and Munn, Jeffrey A. and Nichol, Robert C. and Padmanabhan, Nikhil and Thakar, Aniruddha R. and Weinberg, David H. and York, Donald},
    title = {Stellar masses and star formation histories for 105 galaxies from the Sloan Digital Sky Survey},
    journal = {Monthly Notices of the Royal Astronomical Society},
    volume = {341},
    number = {1},
    pages = {33-53},
    year = {2003},
    month = {05},
    issn = {0035-8711},
    doi = {10.1046/j.1365-8711.2003.06291.x},
    url = {https://doi.org/10.1046/j.1365-8711.2003.06291.x},
    eprint = {https://academic.oup.com/mnras/article-pdf/341/1/33/18650863/341-1-33.pdf},
}

@article{Kriek_2006,
doi = {10.1086/508371},
url = {https://doi.org/10.1086/508371},
year = {2006},
month = {sep},
publisher = {},
volume = {649},
number = {2},
pages = {L71},
author = {Kriek, Mariska and van Dokkum, Pieter G. and Franx, Marijn and Quadri, Ryan and Gawiser, Eric and Herrera, David and Illingworth, Garth D. and Labbé, Ivo and Lira, Paulina and Marchesini, Danilo and Rix, Hans-Walter and Rudnick, Gregory and Taylor, Edward N. and Toft, Sune and Urry, C. Megan and Wuyts, Stijn},
title = {Spectroscopic Identification of Massive Galaxies at z ~ 2.3 with Strongly Suppressed Star Formation*},
journal = {The Astrophysical Journal}
}

@misc{Bluck_2023,
      title={Galaxy quenching at the high redshift frontier: A fundamental test of cosmological models in the early universe with JWST-CEERS}, 
      author={Asa F. L. Bluck and Christopher J. Conselice and Katherine Ormerod and Joanna M. Piotrowska and Nathan Adams and Duncan Austin and Joseph Caruana and K. J. Duncan and Leonardo Ferreira and Paul Goubert and Thomas Harvey and James Trussler and Roberto Maiolino},
      year={2023},
      eprint={2311.02526},
      archivePrefix={arXiv},
      primaryClass={astro-ph.GA},
      url={https://arxiv.org/abs/2311.02526}, 
}

@article{Lerner_2014,
   title={UV surface brightness of galaxies from the local universe to z ~ 5},
   volume={23},
   ISSN={1793-6594},
   url={http://dx.doi.org/10.1142/S0218271814500588},
   DOI={10.1142/s0218271814500588},
   number={06},
   journal={International Journal of Modern Physics D},
   publisher={World Scientific Pub Co Pte Lt},
   author={Lerner, Eric J. and Falomo, Renato and Scarpa, Riccardo},
   year={2014},
   month=May, pages={1450058} }

@article{van_der_Wel_2014,
   title={3D-HST+CANDELS: THE EVOLUTION OF THE GALAXY SIZE-MASS DISTRIBUTION SINCE<i>z</i>= 3},
   volume={788},
   ISSN={1538-4357},
   url={http://dx.doi.org/10.1088/0004-637X/788/1/28},
   DOI={10.1088/0004-637x/788/1/28},
   number={1},
   journal={The Astrophysical Journal},
   publisher={American Astronomical Society},
   author={van der Wel, A. and Franx, M. and van Dokkum, P. G. and Skelton, R. E. and Momcheva, I. G. and Whitaker, K. E. and Brammer, G. B. and Bell, E. F. and Rix, H.-W. and Wuyts, S. and Ferguson, H. C. and Holden, B. P. and Barro, G. and Koekemoer, A. M. and Chang, Yu-Yen and McGrath, E. J. and Häussler, B. and Dekel, A. and Behroozi, P. and Fumagalli, M. and Leja, J. and Lundgren, B. F. and Maseda, M. V. and Nelson, E. J. and Wake, D. A. and Patel, S. G. and Labbé, I. and Faber, S. M. and Grogin, N. A. and Kocevski, D. D.},
   year={2014},
   month=May, pages={28} }

@article{van_Dokkum_2015,
   title={FORMING COMPACT MASSIVE GALAXIES},
   volume={813},
   ISSN={1538-4357},
   url={http://dx.doi.org/10.1088/0004-637X/813/1/23},
   DOI={10.1088/0004-637x/813/1/23},
   number={1},
   journal={The Astrophysical Journal},
   publisher={American Astronomical Society},
   author={van Dokkum, Pieter G. and Nelson, Erica June and Franx, Marijn and Oesch, Pascal and Momcheva, Ivelina and Brammer, Gabriel and Schreiber, Natascha M. Förster and Skelton, Rosalind E. and Whitaker, Katherine E. and Wel, Arjen van der and Bezanson, Rachel and Fumagalli, Mattia and Illingworth, Garth D. and Kriek, Mariska and Leja, Joel and Wuyts, Stijn},
   year={2015},
   month=Oct, pages={23} }

@article{Mowla_2019,
   title={COSMOS-DASH: The Evolution of the Galaxy Size–Mass Relation since z ∼ 3 from New Wide-field WFC3 Imaging Combined with CANDELS/3D-HST},
   volume={880},
   ISSN={1538-4357},
   url={http://dx.doi.org/10.3847/1538-4357/ab290a},
   DOI={10.3847/1538-4357/ab290a},
   number={1},
   journal={The Astrophysical Journal},
   publisher={American Astronomical Society},
   author={Mowla, Lamiya A. and Dokkum, Pieter van and Brammer, Gabriel B. and Momcheva, Ivelina and Wel, Arjen van der and Whitaker, Katherine and Nelson, Erica and Bezanson, Rachel and Muzzin, Adam and Franx, Marijn and MacKenty, John and Leja, Joel and Kriek, Mariska and Marchesini, Danilo},
   year={2019},
   month=July, pages={57} }

@article{Nedkova_2021,
   title={Extending the evolution of the stellar mass–size relation at <i>z</i> ≤ 2 to low stellar mass galaxies from HFF and CANDELS},
   volume={506},
   ISSN={1365-2966},
   url={http://dx.doi.org/10.1093/mnras/stab1744},
   DOI={10.1093/mnras/stab1744},
   number={1},
   journal={Monthly Notices of the Royal Astronomical Society},
   publisher={Oxford University Press (OUP)},
   author={Nedkova, Kalina V and Häußler, Boris and Marchesini, Danilo and Dimauro, Paola and Brammer, Gabriel and Eigenthaler, Paul and Feinstein, Adina D and Ferguson, Henry C and Huertas-Company, Marc and Johnston, Evelyn J and Kado-Fong, Erin and Kartaltepe, Jeyhan S and Labbé, Ivo and Lange-Vagle, Daniel and Martis, Nicholas S and McGrath, Elizabeth J and Muzzin, Adam and Oesch, Pascal and Ordenes-Briceño, Yasna and Puzia, Thomas and Shipley, Heath V and Simmons, Brooke D and Skelton, Rosalind E and Stefanon, Mauro and van der Wel, Arjen and Whitaker, Katherine E},
   year={2021},
   month=June, pages={928–956} }

@misc{Ward_2023,
      title={Evolution of the Size-Mass Relation of Star-forming Galaxies Since $z=5.5$ Revealed by CEERS}, 
      author={Ethan M. Ward and Alexander de la Vega and Bahram Mobasher and Elizabeth J. McGrath and Kartheik G. Iyer and Antonello Calabro and Luca Costantin and Mark Dickinson and Benne W. Holwerda and Marc Huertas-Company and Michaela Hirschmann and Ray A. Lucas and Viraj Pandya and Stephen M. Wilkins and L. Y. Aaron Yung and Pablo Arrabal Haro and Micaela B. Bagley and Steven L. Finkelstein and Jeyhan S. Kartaltepe and Anton M. Koekemoer and Casey Papovich and Nor Pirzkal},
      year={2023},
      eprint={2311.02162},
      archivePrefix={arXiv},
      primaryClass={astro-ph.GA},
      url={https://arxiv.org/abs/2311.02162}, 
}

@ARTICLE{Hodge_2016,
       author = {{Hodge}, J.~A. and {Swinbank}, A.~M. and {Simpson}, J.~M. and {Smail}, I. and {Walter}, F. and {Alexander}, D.~M. and {Bertoldi}, F. and {Biggs}, A.~D. and {Brandt}, W.~N. and {Chapman}, S.~C. and {Chen}, C.~C. and {Coppin}, K.~E.~K. and {Cox}, P. and {Dannerbauer}, H. and {Edge}, A.~C. and {Greve}, T.~R. and {Ivison}, R.~J. and {Karim}, A. and {Knudsen}, K.~K. and {Menten}, K.~M. and {Rix}, H.-W. and {Schinnerer}, E. and {Wardlow}, J.~L. and {Weiss}, A. and {van der Werf}, P.},
        title = "{Kiloparsec-scale Dust Disks in High-redshift Luminous Submillimeter Galaxies}",
      journal = {\apj},
         year = 2016,
        month = dec,
       volume = {833},
       number = {1},
          eid = {103},
        pages = {103},
          doi = {10.3847/1538-4357/833/1/103},
archivePrefix = {arXiv},
       eprint = {1609.09649},
 primaryClass = {astro-ph.GA},
       adsurl = {https://ui.adsabs.harvard.edu/abs/2016ApJ...833..103H}
}

@article{Barro_2016,
   title={SUB-KILOPARSEC ALMA IMAGING OF COMPACT STAR-FORMING GALAXIES AT z ∼ 2.5: REVEALING THE FORMATION OF DENSE GALACTIC CORES IN THE PROGENITORS OF COMPACT QUIESCENT GALAXIES},
   volume={827},
   ISSN={2041-8213},
   url={http://dx.doi.org/10.3847/2041-8205/827/2/L32},
   DOI={10.3847/2041-8205/827/2/l32},
   number={2},
   journal={The Astrophysical Journal Letters},
   publisher={American Astronomical Society},
   author={Barro, G. and Kriek, M. and Pérez-González, P. G. and Trump, J. R. and Koo, D. C. and Faber, S. M. and Dekel, A. and Primack, J. R. and Guo, Y. and Kocevski, D. D. and Muñoz-Mateos, J. C. and Rujoparkarn, W. and Seth, K.},
   year={2016},
   month=Aug, pages={L32} }

@article{G_mez_Guijarro_2022,
   title={GOODS-ALMA 2.0: Source catalog, number counts, and prevailing compact sizes in 1.1 mm galaxies},
   volume={658},
   ISSN={1432-0746},
   url={http://dx.doi.org/10.1051/0004-6361/202141615},
   DOI={10.1051/0004-6361/202141615},
   journal={Astronomy \& Astrophysics},
   publisher={EDP Sciences},
   author={Gómez-Guijarro, C. and Elbaz, D. and Xiao, M. and Béthermin, M. and Franco, M. and Magnelli, B. and Daddi, E. and Dickinson, M. and Demarco, R. and Inami, H. and Rujopakarn, W. and Magdis, G. E. and Shu, X. and Chary, R. and Zhou, L. and Alexander, D. M. and Bournaud, F. and Ciesla, L. and Ferguson, H. C. and Finkelstein, S. L. and Giavalisco, M. and Iono, D. and Juneau, S. and Kartaltepe, J. S. and Lagache, G. and Le Floc’h, E. and Leiton, R. and Lin, L. and Motohara, K. and Mullaney, J. and Okumura, K. and Pannella, M. and Papovich, C. and Pope, A. and Sargent, M. T. and Silverman, J. D. and Treister, E. and Wang, T.},
   year={2022},
   month=Jan, pages={A43} }

@ARTICLE{Tacchella_2019,
       author = {{Tacchella}, Sandro and {Diemer}, Benedikt and {Hernquist}, Lars and {Genel}, Shy and {Marinacci}, Federico and {Nelson}, Dylan and {Pillepich}, Annalisa and {Rodriguez-Gomez}, Vicente and {Sales}, Laura V. and {Springel}, Volker and {Vogelsberger}, Mark},
        title = "{Morphology and star formation in IllustrisTNG: the build-up of spheroids and discs}",
      journal = {\mnras},
         year = 2019,
        month = aug,
       volume = {487},
       number = {4},
        pages = {5416-5440},
          doi = {10.1093/mnras/stz1657},
archivePrefix = {arXiv},
       eprint = {1904.12860},
 primaryClass = {astro-ph.GA},
       adsurl = {https://ui.adsabs.harvard.edu/abs/2019MNRAS.487.5416T}
}

@ARTICLE{Husko_2023,
       author = {{Hu{\v{s}}ko}, Filip and {Lacey}, Cedric G. and {Baugh}, Carlton M.},
        title = "{The buildup of galaxies and their spheroids: The contributions of mergers, disc instabilities, and star formation}",
      journal = {\mnras},
         year = 2023,
        month = feb,
       volume = {518},
       number = {4},
        pages = {5323-5339},
          doi = {10.1093/mnras/stac3152},
archivePrefix = {arXiv},
       eprint = {2207.07139},
 primaryClass = {astro-ph.GA},
       adsurl = {https://ui.adsabs.harvard.edu/abs/2023MNRAS.518.5323H}
}

\end{document}